\documentclass[onecolumn]{els-mrw}             

\usepackage{amsmath,amssymb,amsfonts,amsthm,makeidx,graphicx}
\usepackage{txfonts}
\usepackage{helvet}

\usepackage{lipsum}

\usepackage{bm}

\newcommand{\rr}{\mbox{\boldmath $r$}}
\newcommand{\ee}{\mbox{\boldmath $e$}}
\newcommand{\pp}{\mbox{\boldmath $p$}}
\newcommand{\EE}{\mbox{\boldmath $E$}}
\newcommand{\HH}{\mbox{\boldmath $H$}}
\newcommand{\BB}{\mbox{\boldmath $B$}}
\newcommand{\DD}{\mbox{\boldmath $D$}}

\newcommand{\sE}{\mbox{\boldmath ${E}_s$}}
\newcommand{\sH}{\mbox{\boldmath ${H}_s$}}
\newcommand{\PP}{\mbox{\boldmath $P$}}
\newcommand{\e}{\mbox{\boldmath $e$}}
\begin{document}

\chapter{Chaotic Billiard Lasers}\label{chap1}

\author[1]{Takahisa Harayama}%


\address[1]{\orgname{Waseda University}, 
\orgdiv{Department of Applied Physics, School of Advanced Science and Engineering}, 
\orgaddress{3-4-1 Okubo, Shinjuku-ku, 169-8555, Tokyo, Japan}}

\articletag{Chapter Article tagline: Sept. 1, 2025}

\maketitle

\begin{glossary}[Keywords]
Microcavity Lasers, Quantum Billiards, Optical Billiards, Dynamical Tunneling, Chaos-Assisted Light Emission, Non-Hermitian Physics, Resonances, Maxwell-Bloch Equations.

\end{glossary}

\begin{abstract}[Abstract]
This chapter provides an overview of chaotic billiard lasers as a prominent branch of quantum chaos. These lasers offer an ideal experimental platform for demonstrating the principles of quantum chaos within a physical system. We begin by introducing the fundamental principles of chaotic ray dynamics in optical microcavities, where the transition from regular to fully chaotic dynamics fundamentally alters the underlying wavefunctions and lasing properties. A central focus is placed on ``chaos-assisted light emission," which serves as a practical manifestation of ``chaos-assisted tunneling"—a hallmark phenomenon in the study of quantum chaos. We discuss both theoretical frameworks and experimental validations, demonstrating how chaotic orbits facilitate the coupling between evanescently localized modes and far-field emission.
Furthermore, exploring how the presence of a gain medium influences established results from quantum chaos research remains a fundamental and intriguing problem in physics. To address this, we establish a rigorous and comprehensive derivation of the Maxwell-Bloch equations for two-dimensional microcavity lasers, specifically examining their application to fully chaotic, stadium-shaped billiard lasers. By bridging the gap between nonlinear lasing processes and chaotic wavefunctions, this chapter highlights the unique potential of chaotic billiards for controlling light-matter interactions and shaping the next generation of unconventional coherent light sources.

\end{abstract}

\section{Introduction}
\label{introduction}

The exploration of quantum chaos, or more broadly wave chaos, has emerged as a cornerstone of modern nonlinear physics \cite{Stoeckmann1999, Haake2010, nakamura2004quantum}. At its core, this field seeks to unravel the profound and often counterintuitive ways in which the chaotic nature of a classically deterministic system manifests itself within the corresponding wave-mechanical framework. Billiards—idealized systems where a point particle undergoes specular reflections within a closed, friction-free geometry—serve as the quintessential paradigm for investigating the transition from integrability to ergodicity. Depending on the boundary's curvature and symmetry, these systems span the entire dynamical spectrum, from the predictable trajectories of circular or rectangular tables to the fully chaotic, mixing dynamics \cite{chernov2006chaotic}.

The historical development of billiard dynamics can be traced back to the foundational work of Lord Kelvin, who recognized the complexity of particle motion in confined geometries \cite{kelvin1901nineteenth}. However, the rigorous mathematical formalization of chaos in these systems was realized in the 20th century through the seminal contributions of Sinai \cite{Sinai1970} and Bunimovich \cite{Bunimovich1979}. They demonstrated that even simple geometric shapes, such as a square with a circular interior scatterer (Sinai billiard) or a domain composed of two semicircles connected by straight segments (Bunimovich stadium), can exhibit ergodicity and mixing properties, where any infinitesimal deviation in initial conditions leads to an exponential divergence of trajectories.

The transition from classical to quantum mechanics within these domains gave birth to the study of quantum billiards \cite{casati1980connection, berry1981quantizing}. The stationary states of a quantum particle are governed by the Helmholtz equation, and the resulting energy spectra reveal the "signature of chaos." In a breakthrough that is one of the most celebrated achievements in quantum chaos, the Bohigas-Giannoni-Schmit (BGS) conjecture proposed that in the semiclassical regime, where the action is much larger than the Planck constant, the spectral fluctuations of chaotic quantum systems are universal and well described by random matrix theory, specifically the Gaussian Orthogonal Ensemble (GOE) \cite{Bohigas1984}.

The implications of this universality extend beyond closed systems to open quantum billiards with channels. For instance, the predicted manifestations of chaos in the magnetoconductance fluctuations of such open cavities have been rigorously confirmed through experimental observations \cite{jalabert1990conductance, marcus1992conductance, nakamura2004quantum}.
 
Parallel to these developments, the field of wave chaos expanded into classical wave systems, notably microwave cavities \cite{stockmann1990quantum, bittner2012pt, Stoeckmann1999} and optical microcavities \cite{nockel1994q, Noeckel1997, Gmachl1998, lee2002observation, harayama2003stable, Fukushima2004, lee2004quasiscarred, lebental2006highly, Wiersig2008, song2009chaotic}. These systems have long served as fundamental platforms for exploring light-matter interactions, ranging from cavity quantum electrodynamics to the development of compact coherent light sources. In recent decades, the study of these microcavities has been profoundly enriched by the insights of quantum chaos. While early laser designs primarily relied on highly symmetric geometries, such as circular or rectangular cavities, the introduction of broken symmetries and deformed boundaries has opened a new frontier: optical billiards and chaotic billiard lasers. The advent of these systems has significantly expanded the horizons of the field \cite{harayama2011two, cao2015dielectric}.

Optical billiards are dielectric microstructures where light is confined inside a billiard cavity. However, unlike the idealized quantum billiards, which are typically Hermitian systems with infinite potential barriers (Dirichlet or Neumann conditions) and real eigenvalues, optical billiards are inherently open and non-Hermitian \cite{schomerus2000quantum, lee2008divergent, wiersig2008asymmetric, lee2009observation, wiersig2011nonorthogonal, cao2015dielectric}. The light field is partially confined by the refractive index contrast between the cavity and its surroundings, leading to the emergence of resonances with complex eigenfrequencies. In these open systems, the imaginary part of the frequency governs the radiation loss, which corresponds to the inverse of the photon lifetime within the cavity, while the real part corresponds to the oscillation frequency, necessitating a description based on outgoing wave conditions at infinity. The refractive index mismatch at the cavity interface allows for radiative emission and evanescent leakage. Consequently, the challenge in optical billiards lies in understanding how the chaotic ray dynamics in the phase space—characterized by stretching and folding processes, stable/unstable manifolds, chaotic seas, and stability islands—dictate the Quality (Q) factors and the highly directional emission patterns of these open resonances.

Research on chaotic billiard lasers—which incorporate an active medium within an optical billiard—is a prominent branch of quantum chaos that explores how the transition from regular to fully chaotic ray dynamics fundamentally reshapes the underlying wave functions and lasing properties. Unlike their integrable counterparts, where modes are well-defined by conserved quantities in ray dynamics, chaotic resonators exhibit complex modal structures governed by the statistical properties of chaotic dynamics. One of the most fascinating phenomena found in quantum chaos is ``chaos-assisted tunneling," which provides a bridge between classically disjoint regions of phase space \cite{bohigas1993manifestations, tomsovic1994chaos, steck2001observation}. In the context of chaotic billiard lasers, this mechanism manifests itself as ``chaos-assisted light emission", offering a unique pathway for coupling evanescently localized energy into the far-field with high directional control \cite{shinohara2010, shinohara2011chaos}. In ``mixed" systems, where stability islands coexist with a chaotic sea, light confinement and emission can be achieved by total internal reflection for stability islands and dynamical tunneling from stability islands to a chaotic sea, respectively. 
The phase space description of ray and wave dynamics and the semiclassical treatment of chaos-assisted tunneling are provided by Martina Hentschel and Akira Shudo, respectively, in this Encyclopedia.  

Historically, the motivation for studying chaotic billiard lasers arose from the need for low-threshold directional emission from microcavity devices \cite{Noeckel1997, Gmachl1998, Wiersig2008}. While conventional whispering-gallery-mode (WGM) lasers based on circular geometries offer high Q factors, they suffer from isotropic emission and poor output coupling. By intentionally deforming the cavity shape, researchers introduced chaos into the ray dynamics to break the symmetry. This also led to the discovery of chaos-assisted light emission, where light localized in stable regions of phase space escapes through the surrounding chaotic sea by dynamical tunneling and dynamical eclipsing, achieving directional emission due to phase space structures \cite{shinohara2010, shinohara2011chaos}. 
To validate these theoretical frameworks, we also present detailed experimental results on the oscillation characteristics of chaotic billiard lasers, particularly the experimental verification of chaos-assisted light emission in deformed microcavities. 
By comparing experimental near-field and far-field patterns with both ray-dynamical and resonance-mode numerical analyses, we confirm that chaotic dynamics can be harnessed to achieve controllable coherent light emission. 

Despite the successes of ray-dynamical and wave-optical descriptions, a complete understanding of chaotic billiard lasers requires a rigorous treatment of the nonlinear interactions between the electromagnetic field and the active gain medium. Instead of a simple linear superposition, the electromagnetic field and the atomic population of the medium are coupled in a self-consistent, nonlinear feedback loop. The medium provides gain to compensate for the radiative losses of the chaotic cavity, while the field induces saturation and spatial hole burning, which in turn modifies the gain available to other modes. The stationary lasing states of such a device are fundamentally non-equilibrium steady states, corresponding to the stable fixed points or attractors of the high-dimensional Maxwell-Bloch equations \cite{harayama1999nonlinear, harayama2003stable, harayama2005theory, Tureci2006}. One of the most striking manifestations of this nonlinearity in fully chaotic systems is the spontaneous symmetry breaking of lasing patterns, where the final lasing state can manifest itself as an asymmetric intensity distribution despite the geometric symmetry of the cavity \cite{Harayama2003, harayama2005theory}. Furthermore, these lasers exhibit a remarkable tendency toward universal single-mode lasing, suggesting a profound self-organizing property \cite{sunada2013stable, sunada2016signature, harayama2017universal, you2022universal}. 

This chapter provides an overview of the physics and applications of chaotic billiard lasers. We begin with a rigorous derivation of the wave equations for two-dimensional optical confinement, considering both the bulk and thin-limit regimes. We then delve into the phase-space representation of ray dynamics using Birkhoff coordinates, establishing the ray-wave correspondence that governs light emission. Special attention is given to the transition from partially chaotic (mixed) systems to fully chaotic stadium billiards, where the breakdown of the conventional critical-angle condition necessitates a nonlinear theoretical framework. This suggests that the conventional ``critical angle" criterion, which is so successful in circular or deformed microdisks, is insufficient to explain the stable lasing observed in fully chaotic semiconductor lasers \cite{Fukushima2004}. To accurately describe how the gain medium forms a stable mode, a rigorous nonlinear description that accounts for all orders of field-medium interaction is essential.
We explore the transition from the linear regime near the threshold—where we identify the frequency pulling effect and the stabilization of resonant modes—to the fully nonlinear regime through intensive dynamical simulations. By comparing numerical results from the Finite-Difference Time-Domain (FDTD) method with those from the Boundary Element Method (BEM), we investigate the stability and spatial structure of lasing modes in the Bunimovich stadium. 
Our results highlight the profound interplay between wave chaos and nonlinear dissipation. By bridging nonlinear dynamics, quantum chaos, and laser physics, chaotic billiard lasers not only serve as a laboratory for fundamental science, but also offer a novel paradigm for the design of next-generation photonic sources.

\section{Optical Billiards}

\subsection{Billiards and Quantum Billiards}

From the perspective of nonlinear physics, billiard systems offer three distinct advantages. First, they provide a rigorous mathematical foundation; certain shapes and configurations allow for analytical proofs that remain elusive in more complex Hamiltonian systems. Second, the trajectories within billiard tables—in both configuration and phase space—offer intuitive visualizations that can be readily mapped onto other physical models. Third, billiard systems have been experimentally realized across diverse platforms, including microwave, semiconductor, droplets, jets, and polymer-based devices. This versatility enables not only the fundamental verification of theoretical predictions but also practical applications such as unconventional optical sources and photonic sensors.

A classical billiard is defined by a point particle moving frictionlessly on a table, reflecting off the boundaries according to the law of reflection (specular reflection). These systems serve as ideal models for studying chaos in classical dynamical systems \cite{chernov2006chaotic}.

When a quantum particle is confined within such a boundary, the system is termed a quantum billiard, a cornerstone model for the study of quantum chaos \cite{Stoeckmann1999, Haake2010, nakamura2004quantum, casati1980connection, berry1981quantizing, Bohigas1984, heller1984bound, mcdonald1979spectrum, mcdonald1988wave, robnik1983classical, gaspard1989semiclassical, gaspard1989exact, sieber1990classical, jalabert1990conductance, dietz1993scattering, tasaki1997interior, backer1998rate}. The stationary states of a quantum particle in this context are governed by the Helmholtz equation, which also describes the stationary modes of classical electromagnetic fields confined within a cavity. Research in quantum chaos has revealed how classical chaos manifests in quantum phenomena within the semiclassical regime—where the action is much larger than the Planck constant. Crucially, these manifestations rely on the wave nature of the particle. Consequently, the signature of classical chaos in wave phenomena has been extensively studied as wave chaos, using microwaves and light trapped within billiard-shaped boundaries, commonly referred to as microwave billiards and optical billiards, respectively \cite{stockmann1990quantum, bittner2012pt, Stoeckmann1999, Noeckel1997, Gmachl1998, cao2015dielectric, harayama2011two}.

\subsection{Optical Billiards}

In contrast to a quantum particle, which is perfectly confined within a billiard, the light field in an optical billiard is only partially confined due to the refractive index contrast between the cavity and its surroundings. Consequently, a portion of the light inevitably leaks or emits from the cavity. This physical difference dictates the boundary conditions: while quantum billiards typically impose Dirichlet or Neumann conditions, optical billiards require the outgoing wave condition. 

As a result, the eigenfrequencies of the Helmholtz equation for optical billiards are complex, where the real and imaginary parts correspond to the oscillation frequency and the decay rate of the eigenmode, respectively. This stands in sharp contrast to the real-valued eigenenergies found in quantum billiards. Mathematically, this implies that optical billiards are fundamentally non-Hermitian systems, distinguishing them from the Hermitian nature of traditional quantum billiards \cite{cao2015dielectric}.

\subsection{Chaotic Billiard Lasers}

Optical billiards that incorporate an active gain medium are referred to as billiard lasers. A fundamental distinction exists between these and their passive counterparts: in quantum, microwave, and optical billiards, the wave functions (or modes) do not interact with one another. In contrast, in a billiard laser, the modes are coupled through their interaction with the active lasing medium. Consequently, while quantum, microwave, and optical billiards are inherently linear systems, billiard lasers are nonlinear systems. Within this framework, stationary lasing states are understood as non-equilibrium steady states corresponding to the stable fixed points of nonlinear time-evolution equations \cite{harayama2003stable, Harayama2003, harayama2005theory, harayama2011two}. 

When the underlying ray dynamics of these systems exhibit chaos, they are termed chaotic billiard lasers \cite{stone2010chaotic}. Such lasers have been experimentally realized using two-dimensional microcavities fabricated from various materials, including semiconductors, polymers, liquid droplets, and micro-jets \cite{nockel1994q, Gmachl1998, lee2002observation,  Fukushima2004, lebental2006highly, song2009chaotic}.



\section{Electromagnetic waves in optical billiards}
\subsection{Maxwell equations and wave equations}
\label{MB eqs}

Let us consider a dielectric material containing a lasing medium, where the refractive index changes abruptly at the material boundary. In such a system, the light field and the atomic states within the cavity are strongly coupled; the field influences the atoms, which in turn provide feedback that modifies the field. This nonlinear interaction between the light field and the medium, in conjunction with the boundary conditions, ultimately determines the stationary lasing modes 
\cite{harayama2003stable, Harayama2003, harayama2005theory, harayama2011two}.

The electro-magnetic fields $\EE$ and $\HH$ are governed by Maxwell's equations  
incorporating the polarization field $\PP$, which represents 
the nonlinear response of the lasing medium,

\begin{equation}
\nabla \times \EE = -\frac{1}{c}
\frac{\partial \BB}{\partial t},           
\label{Maxwell1}
\end{equation}
\begin{equation}
\nabla \cdot \DD = 0 ,
\label{Maxwell2}
\end{equation}
\begin{equation}
\nabla \times \HH = \frac{1}{c}
\frac{\partial \DD}{\partial t}, 
\label{Maxwell3}
\end{equation}
\begin{equation}
\nabla \cdot \BB = 0,                       
\label{Maxwell4}
\end{equation}
where 
\begin{equation}
\DD = \epsilon \EE + 4 \pi \PP,       
\label{Maxwell5}
\end{equation}
\begin{equation}
\BB = \mu \HH. 
\label{Maxwell6}
\end{equation}
Here,  $\DD$ is the electric displacement and $\BB$ the magnetic induction.

In this chapter, the dielectric material is structured as an optical billiard, whose boundaries are assumed to be strictly vertical along the $z$-axis.  
Two extreme cases are considered depending on the vertical extent $2d$ of the optical confinement region relative to the wavelength $\lambda$.

In the bulk regime ($2d\gg \lambda$), where the vertical dimension is sufficiently large, 
the light field can be treated as propagating in a bulk medium. 
Conversely, in the thin-limit regime ($2d\ll \lambda$), which corresponds to structures such as single quantum well (SQW) lasers, the light is strongly confined within a sub-wavelength layer. While the $z$-dependence of the fields may be neglected in both cases for different physical reasons, 
the resulting effective refractive indices differ significantly. 
Therefore, when comparing theoretical results with experimental data, the precise determination of the effective refractive index is of critical importance.

For the above two extreme cases, by neglecting the $z$-dependence of the electromagnetic fields—the physical justification for which will be detailed in the following subsection—Eqs.~(\ref{Maxwell1}) and (\ref{Maxwell3}) reduce to the following components,

\begin{equation}
\frac{\partial E_z}{\partial y} = -\frac{\mu}{c} 
\frac{\partial H_x}{\partial t},  
\ \ 
-\frac{\partial E_z}{\partial x}
= -\frac{\mu}{c} \frac{\partial H_y}{\partial t},  
\ \ 
\frac{\partial E_y}{\partial x} -\frac{\partial E_x}{\partial y}
= -\frac{\mu}{c} \frac{\partial H_z}{\partial t},  
\label{MaxwellComp1}
\end{equation}
\begin{equation}
\frac{\partial H_z}{\partial y} = \frac{1}{c} 
\frac{\partial}{\partial t}(\epsilon E_x +4\pi P_x),  
\ \ 
-\frac{\partial H_z}{\partial x}
= \frac{1}{c} 
\frac{\partial}{\partial t}(\epsilon E_y +4\pi P_y),  
\ \ 
\frac{\partial H_y}{\partial x} -\frac{\partial H_x}{\partial y}
= \frac{1}{c} 
\frac{\partial}{\partial t}(\epsilon E_z +4\pi P_z).  
\label{MaxwellComp2}
\end{equation}
In these equations, if $H_z(E_z)$ vanishes, then $E_x(H_x)$ and $E_y(H_y)$ must also be zero.  
Consequently, the solutions to Eqs.~(\ref{MaxwellComp1}) and (\ref{MaxwellComp2}) decouple into transverse magnetic (TM) modes ($H_z=0$) and transverse electric (TE) modes ($E_z=0$), 
leading to the following wave equations,
\begin{equation}
\left(\nabla_{x y}^{2}-\frac{n^2}{c^2} \frac{{\partial}^2}{\partial t^2} \right)E_z=\frac{4\pi\mu}{c^2}\frac{\partial^2P_z\ }{\partial t^2},
\label{TMdynamics}
\end{equation}
for TM modes, and 
\begin{equation}
\left(\nabla_{x y}^{2}-\frac{n^2}{c^2} \frac{{\partial}^2}{\partial t^2} \right) H_z=-\frac{4\pi }{c}\frac{\partial}{\partial t}\left(\frac{\partial P_y}{\partial x}-\frac{\partial P_x}{\partial y}\right), 
\label{TEdynamics}
\end{equation}
for TE modes. In the above, $n$ denotes the refractive index and equals 
$\sqrt{\epsilon \mu}$.
Since the dielectric medium is non-magnetic, the magnetic permeability $\mu$ is taken to be unity throughout the following derivations.

\subsection{Effective refractive indices in two-dimensional Helmholtz equations}
\label{thin film}


In the previous subsection, we neglected the $z$-dependence of the electromagnetic fields in the two extreme cases:  
the bulk-limit ($2d\gg \lambda$) and the thin-limit ($2d\ll \lambda$). 
Here, we provide the physical justification for this approximation and derive the corresponding effective refractive indices for each regime.

Consider an electromagnetic field oscillating at a frequency $\omega$: 
$\EE(\rr,t)=\sE(\rr) \cos{\omega t}$ and $\HH(\rr,t)=\sH(\rr) \cos{\omega t}$. 
For simplicity, we first neglect the effects of the lasing medium. The stationary wave equations are given by
\begin{equation} 
\left(
\frac{\partial^2}{\partial x^2}+\frac{\partial^2}{\partial y^2}
+\frac{\partial^2}{\partial z^2}
\right) 
\begin{pmatrix} \sE \\ \sH \end{pmatrix}
= -n^2k^2 \begin{pmatrix} \sE \\ \sH \end{pmatrix},
\end{equation}
where $k=\omega/c$. The cavity we consider here is translationally invariant along the $z$-axis same as before. 
If the thickness of the device is much larger than the wavelength, the fields can be assumed to be independent of $z$.
In this case, we obtain the standard two-dimensional Helmholtz equation with the bulk refractive index $n$.

In the following, we assume that the vertical extent $2d$ of the optical confinement region is much smaller than the wavelength.
Assuming that the electromagnetic fields can be factorized as $h^{E(H)}_{x,y,z}(z)f^{E(H)}_{x,y,z}(x,y)$, 
we obtain the following separated equation: 
\begin{equation}
\frac{1}{f^{E(H)}_{x,y,z}(x,y)}
\left(
\frac{\partial^2}{\partial x^2}+\frac{\partial^2}{\partial y^2}
\right)
f^{E(H)}_{x,y,z}(x,y)+
\frac{1}{h^{E(H)}_{x,y,z}(z)}\frac{\partial^2}{\partial z^2} 
h^{E(H)}_{x,y,z}(z)
=-n^2k^2.
\end{equation}
This leads to the two-dimensional Helmholtz equation
\begin{equation}
\left(
\frac{\partial^2}{\partial x^2}+\frac{\partial^2}{\partial y^2}
\right)
f^{E(H)}_{x,y,z}(x,y)= -\beta(k)^2 f^{E(H)}_{x,y,z}(x,y),
\label{Thinfilm}
\end{equation}
and the transverse equation
\begin{equation}
\frac{\partial^2}{\partial z^2} h^{E(H)}_{x,y,z}(z)
=(-n^2k^2+\beta(k)^2) h^{E(H)}_{x,y,z}(z),
\label{h(z)}
\end{equation}
where $\beta(k)$ is a propagation constant which depends on $k$. 
For example, from Maxwell's equations, the continuity of the tangential components $E_{x,y}$ at the interfaces ($z=\pm d$) 
must be satisfied. Accordingly, the solutions to Eq.~(\ref{h(z)}) take the form  
\begin{equation}
h_{x,y}(z) =
 \left\{
 \begin{array}{ll}
   \left( A \cos (\kappa d) + B \sin (\kappa d) \right) 
  e^{- \gamma \left( z - d \right)},
  & \quad \mbox{ for $ z \geq d $}  \\
  A \cos (\kappa z) + B \sin (\kappa z), 
  & \quad \mbox{ for $ -d\leq z < d $} \\
   \left( A \cos (\kappa d) - B \sin (\kappa d) \right) 
  e^{ \gamma \left( z + d \right)},
  & \quad \mbox{ for $ z < -d $}  
 \end{array}
 \right.
\label{field}
\end{equation}
where 
$
\kappa^2 = n_{in}^2k^2-\beta(k)^2
$ and 
$
\gamma^2 = -n_{out}^2k^2+\beta(k)^2.
$
Then, the continuity of the normal derivative of $E_{x,y}$ at the interfaces leads to the transcendental equation
\begin{equation}
\tan (2\kappa d) = \frac{2 \kappa \sqrt{q^2 - \kappa^2}}
{2 \kappa^2 - q^2},
\label{bound}
\end{equation}
where
\begin{equation}
q^2 =  \left(n_{in}^2 - n_{out}^2\right) k^2.
\end{equation}
In the limit of a very thin vertical extent $(2d \ll \lambda)$, Eq.~(\ref{bound}) possesses only one bound-state solution for $\kappa$ 
which lies slightly below $q$. 
Consequently, we obtain the two-dimensional Helmholtz equation
\begin{equation}
\left(
\frac{\partial^2}{\partial x^2}+\frac{\partial^2}{\partial y^2}
\right)
f^{E}_{x,y}(x,y)
+n_{\text{eff}}^2 k^2 f^{E}_{x,y}(x,y)=0 ,
\label{2dHelm}
\end{equation}
where the effective refractive index $n_{\text{eff}}=\beta(k)/k$. $h^{E}_{x,y}(z)$ is determined solely by the vertical confinement structure. Since $2d\ll \lambda$, only the fundamental mode is supported for $h^{E}_{x,y}(z)$, allowing the system to be effectively described by the two-dimensional Helmholtz equation (\ref{2dHelm}). For optical billiards, it is therefore sufficient to employ Eq.~(\ref{2dHelm}) for stationary states, 
or Eqs.~(\ref{TMdynamics}) and (\ref{TEdynamics}) for dynamics, by replacing the refractive index $n$ with $n_{\text{eff}}$.
In the subsequent sections, we assume that the refractive index inside the billiard is $n_{\text{in}}=3.3$—a typical value of the effective refractive index for semiconductor lasers—while the surrounding air is taken as $n_{\text{out}}=1$. 


\subsection{Resonances}

In the previous sections, we derived the fundamental wave equations describing electromagnetic fields in optical billiards. Real billiard lasers contain an active medium that interacts with the electromagnetic fields through the polarization $\PP$. The fields evolve according to Maxwell's equations, thereby altering the states of the active medium; conversely, these changes in the medium modify the fields via the polarization term. This nonlinear feedback loop eventually leads to the formation of stable, self-consistent lasing states.

In general, accounting for the full effect of the active lasing medium is crucial for a complete understanding of billiard lasers. However, many essential characteristics—such as the spatial emission patterns and threshold properties—can be effectively captured by the linear eigenmodes of the system. These modes are the resonance solutions of the two-dimensional Helmholtz equation (\ref{2dHelm}). 

Here it is important to note that this result was obtained under the assumption of an electromagnetic field oscillating at a frequency $\omega$ for simplicity: $\EE(\rr,t)=\sE(\rr) \cos{\omega t}$ and $\HH(\rr,t)=\sH(\rr) \cos{\omega t}$. 
However, unless the edge mirrors of the optical billiard are perfect, light is always emitted to the outside. Accordingly, strictly speaking, such decaying or growing states cannot be represented as purely stationary solutions. Consequently, the time dependence of the electromagnetic fields is generalized by allowing the frequencies and amplitudes to be complex:
\begin{equation}
E_z(H_z)=\psi (x,y) e^{-i\omega t}+\psi^* (x,y) e^{i\omega^* t}.
\end{equation}
By substituting this into the dynamical equations (\ref{TMdynamics}) and (\ref{TEdynamics}) and applying the rotating wave approximation (RWA)—which omits the fast-oscillating terms ($e^{\pm i2Re(\omega) t}$)—we obtain the complex two-dimensional Helmholtz equation
\begin{equation}
\left(\nabla_{xy}^2 + n^2 \frac{\omega^2}{c^2}  \right)\psi(x,y)=0, 
\end{equation}
subject to the outgoing wave condition at infinity. 
The boundary conditions at the billiard boundary are given by the continuity of the fields and their derivatives. 
Note that for the TE mode, the normal derivative of the magnetic field $H_z$ is discontinuous at the billiard boundary, satisfying 
$
1/n^2_\text{in}{\partial \psi_\text{in}}/{\partial n}={1}/{n^2_\text{out}}{\partial \psi_\text{out}}/{\partial n}, 
$
where $\partial/\partial n$ denotes the normal derivative at the billiard boundary. 
In this framework, both the eigenfrequency $\omega$ and the eigenfunction $\psi$ are generally complex. 

The complex eigenfrequency $\omega=\omega_r+i\omega_i$ reflects the non-conservative nature of the optical billiard system. 
The real part, $Re(\omega)=\omega_r$, represents the oscillation frequency of the electromagnetic field. 
In contrast, the imaginary part, $Im(\omega)=\omega_i$, characterizes the energy decay rate.

Specifically, the time-averaged electromagnetic energy density $\bar{u}$ within the cavity evolves as $\bar{u}\propto e^{2\omega_i t}$. 
Since the light always leaks out from the billiard edge, the energy decreases over time, which implies $\omega_i < 0$. 
This characterizes the resonances with a finite lifetime $\tau$ defined as
\begin{equation}
\tau \equiv -\frac{1}{2\omega_i}.
\end{equation}
The quality of the optical confinement is typically quantified by the $Q$-factor, $Q\equiv-\omega_r/(2\omega_i)=\omega_r \tau$, where a larger $Q$ indicates a longer-lived resonance with lower radiation loss.
If the $Q$-factor of a resonance is sufficiently high, the minimal radiation loss can be readily compensated by the gain from the active lasing medium, leading to a lower lasing threshold.

\section{Chaos assisted light emission }

\subsection{Birkhoff coordinates and critical line}

In the previous sections, light has been described as a classical electromagnetic wave. In this section, we transition to the ray dynamical properties of light. In the limit where the wavelength is sufficiently small compared to the cavity dimensions (the short-wavelength limit), the wave phenomena can be effectively analyzed through the dynamics of rays. This approach allows us to understand the resonance and lasing characteristics from the perspective of Hamiltonian dynamics and the geometry of the optical billiard.

\begin{figure}[b]
\centering
\includegraphics[width=0.7\columnwidth]{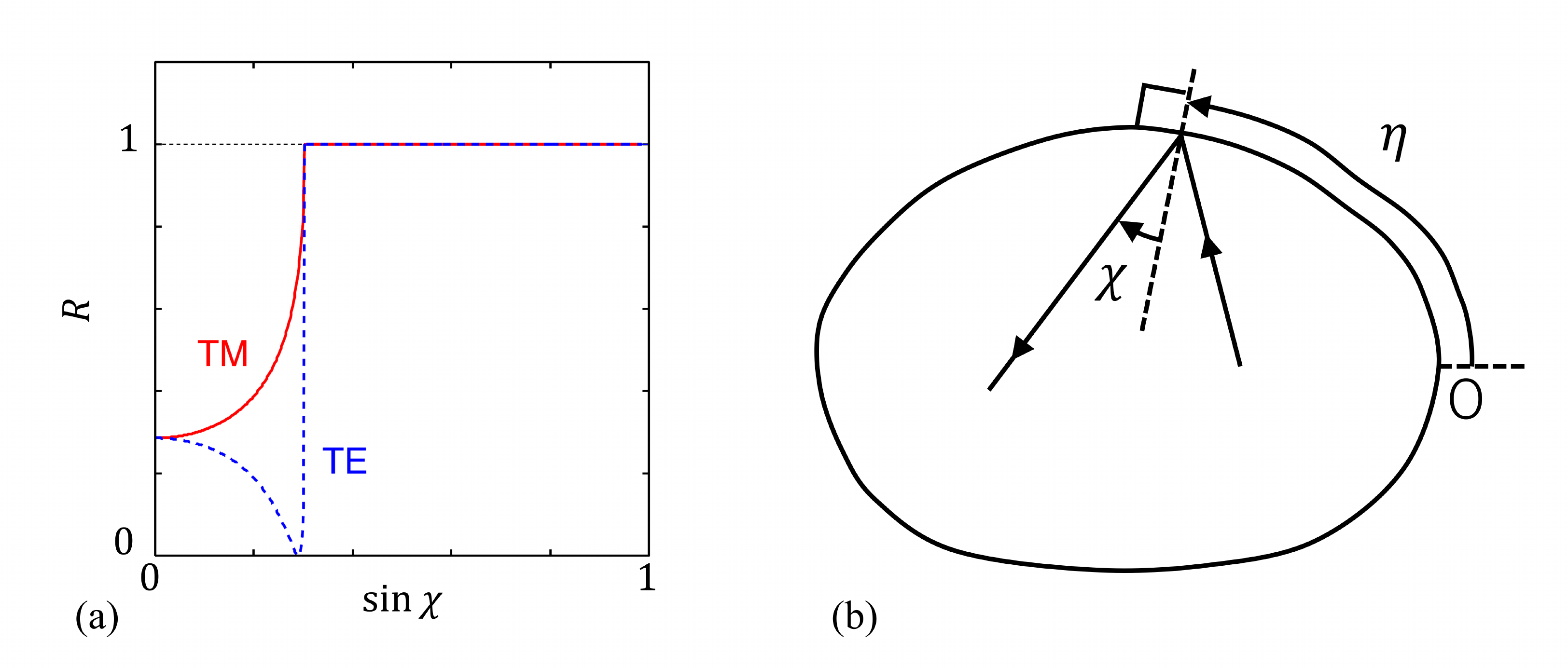}
\caption{
Critical angle and the Birkhoff coordinates. 
(a) Reflectance $R$ as a function of the incident angle $\chi$. 
The solid red and dashed blue curves represent $R_{\text{TM}}$ and $R_{\text{TE}}$, respectively. 
The incident angle where $R_{\text{TE}}$ vanishes is called the Brewster angle.
(b) Schematic illustration of the Birkhoff coordinates $(\eta,\sin \chi)$. 
The reflection angle $\chi$ is equal to the incident angle. Assuming unit mass, the tangential momentum is given by  $\sin \chi$. This serves as the canonical momentum conjugate to the coordinate $\eta$, which represents the arc length along the boundary measured from a fixed origin O. Note that $\eta$ is normalized to unity for a full circumference.
}
\label{fig:CriticalBirk}
\end{figure}

The simplest theoretical criterion for the confinement of light in an optical billiard is the critical angle condition for ray orbits. 
For the TM mode, the reflectance $R_{\text{TM}}$ at an incident angle $\chi$ 
is defined as the ratio of the reflected light intensity to the incident light intensity 
and is given by the square of the Fresnel reflection coefficient
\begin{equation}
R_{\text{TM}}=\left( 
\frac{n_{\text{in}}{\cos\chi-n_{\text{out}}\cos\chi_t}}{n_{\text{in}}\cos\chi+n_{\text{out}}{\cos\chi_t}}
\right)^2.
\label{TMfresnel}
\end{equation}
For the TE mode, the reflectance $R_{\text{TE}}$ is expressed as
\begin{equation}
R_{\text{TE}}=\left(
\frac{n_{\text{in}}\cos\chi_t - n_{\text{out}}\cos\chi}{n_{\text{in}}\cos\chi_t + n_{\text{out}}\cos\chi}
\right)^2.
\label{R_TE}
\end{equation}
In these expressions, $\chi_t$ denotes the angle of refraction, which is related to the incident angle and 
the refractive indices ($n_{\text{in}}$ and $n_{\text{out}}$, respectively,inside and outside the optical billiard) through Snell's law:
\begin{equation}
n_{\text{in}} \sin \chi = n_{\text{out}} \sin \chi_t.
\label{Snell}
\end{equation}
The dependence of reflectances $R_{\text{TM}}$ and $R_{\text{TE}}$ on the incident angle is shown in Fig.~\ref{fig:CriticalBirk}(a). 
The critical angle $\chi_c$ is defined as the incident angle 
for which the angle of refraction $\chi_t$ becomes exactly $\pi/2$. 
At this point, the refracted ray propagates parallel to the billiard edge, 
and for any incident angle $\chi \geq \chi_c$, total internal reflection (TIR) occurs, 
resulting in unity reflectance, i.e., $R=1$. According to Snell's law, the critical angle is expressed as
\begin{equation}
\sin \chi_c = \frac{n_{\text{out}}}{n_{\text{in}}}.
\label{Fresnel}
\end{equation}


Here, Birkhoff coordinates are introduced to parameterize the Poincaré surface of section for billiards, which is convenient not only to describe ray orbits as orbits of a point particle in the phase space of a classical dynamical system but also to confirm the critical angle condition for ray orbits. 
Let us suppose that a ray hits the boundary of the billiard as shown in Fig.~\ref{fig:CriticalBirk}(b).
The tangential momentum of a particle at the reflection point on a billiard boundary is the canonical momentum conjugate to the arc length measured along the boundary. Every time the particle hits the boundary, we record the sine of the incident angle $\chi$ and the arc length $\eta$ measured from a reference origin O, which is normalized to unity for a full circumference. 
This defines a symplectic mapping from $(\eta_n, \sin \chi_n)$ to $(\eta_{n+1}, \sin \chi_{n+1})$, which is area-preserving. 

\begin{figure}[t]
\centering
\includegraphics[width=0.8\columnwidth]{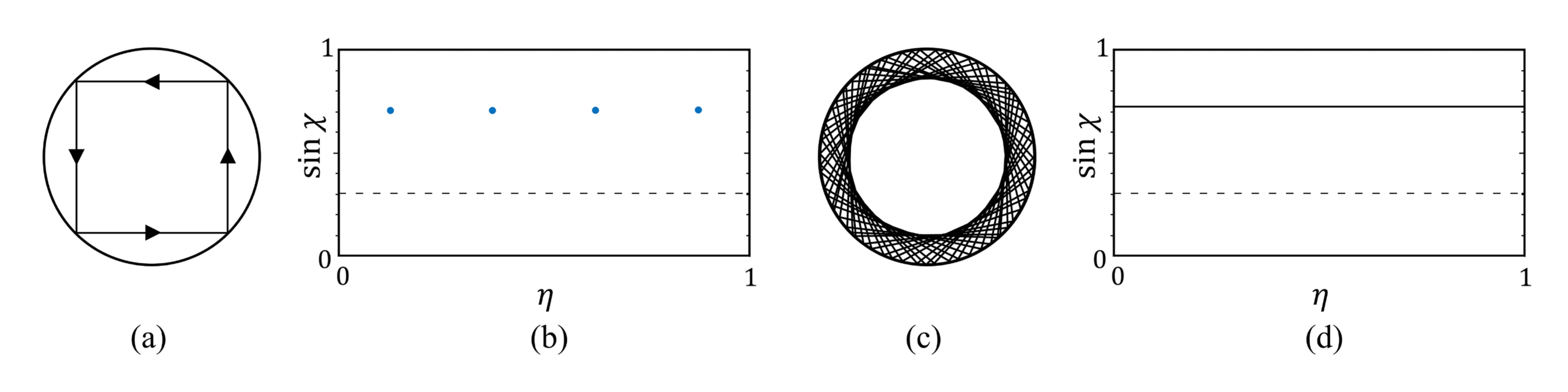}
\caption{
A circular optical billiard. 
(a) A square periodic orbit in the circular billiard with refractive index $n_{\text{in}}=3.3$ 
(assuming $n_{\text{out}}=1$ for the surrounding air). (b) The corresponding representation in the Birkhoff coordinates (the blue dots). The dashed black line indicates the critical line $\sin \chi=\sin \chi_c$ for total internal reflection. 
(c) A trajectory with an incident angle slightly larger than $\pi/4$. 
Due to the irrationality of the angle, the ray never returns to the initial point, filling the boundary densely. 
(d) The corresponding representation in the Birkhoff coordinates. The orbit forms a horizontal invariant line.
}
\label{fig:Circ4PO}
\end{figure}

First, let us consider a circular optical billiard as the simplest example. 
A counterclockwise rotating (CCW) square periodic ray orbit, which undergoes four reflections at the boundary (Fig. \ref{fig:Circ4PO}(a)), is represented in the Birkhoff coordinates as shown in Fig. \ref{fig:Circ4PO}(b).
Since the incident angle of this periodic ray orbit is $45^{\circ}$ at the billiard boundary, which is greater than the critical angle $\chi_c=\arcsin {1/3.3} \approx 17.6^{\circ}$, the CCW square periodic ray orbit is sustained by total internal reflection and is perfectly confined within the circular optical billiard.

When the incident angle is an irrational multiple of $\pi$ slightly larger than $\pi/4$, 
the ray orbit never returns to the initial incident point. 
Since the incident angle remains constant at each reflection, as shown in Fig.~\ref{fig:Circ4PO}(c), 
the orbit corresponds to the horizontal line in the Birkhoff coordinates (Fig.~\ref{fig:Circ4PO}(d)). 
This orbit is perfectly confined within the circular optical billiard by total internal reflection.

\begin{figure}[b]
\centering
\includegraphics[width=0.9\columnwidth]{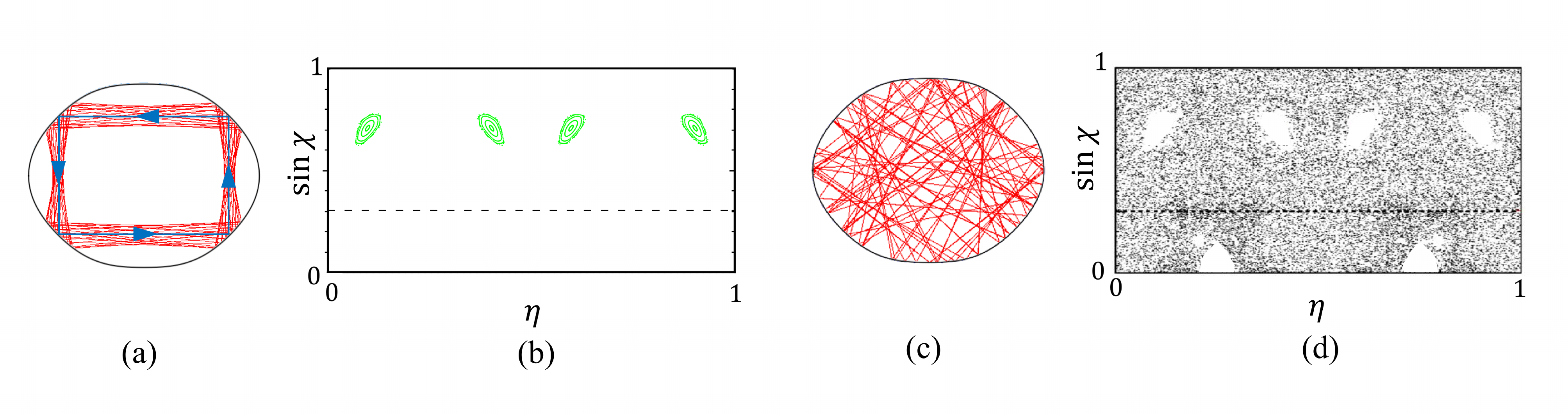}
\caption{
The deformed optical billiard designed by Prof. Evgenii Narimanov. 
(a) The CCW rectangular periodic ray orbit (blue lines) and a neighboring quasi-periodic ray orbit (red lines) in real space. 
(b) The corresponding representation in the Birkhoff coordinates. The periodic points (corresponding to the blue lines in (a), which are not depicted in this figure) are surrounded by a set of four concentric invariant curves (green), each representing a distinct quasi-periodic ray orbit starting from different initial conditions. Collectively, these invariant curves form the stability islands. The dashed black line indicates the critical line $\sin \chi=\sin \chi_c$ for total internal reflection. 
(c) A chaotic ray orbit in real space. Starting from the chaotic region, the ray wanders stochastically and fills the interior of the cavity. 
(d) The corresponding representation in the Birkhoff coordinates. The scattered points form a large chaotic sea, which extends to the critical line $\sin \chi_c$ (dashed line), indicating that these orbits can escape the billiard. The blank regions correspond to the stability islands surrounding the periodic orbits with 4 and 2 reflections, which act as dynamical barriers that chaotic trajectories cannot penetrate.
}
\label{fig:CAT4PO}
\end{figure}

\subsection{Chaotic optical billiard and chaos assisted tunneling light emission}

Next, let us consider an optical billiard deformed from a circle as designed by Prof. Evgenii Narimanov (Purdue University), whose shape is
defined in the polar coordinates $(r,\phi)$ as
\begin{equation}
r=R(\phi)=r_0 (1+a\cos 2\phi+b\cos 4\phi+c\cos 6\phi),
\label{eq:cavity}
\end{equation}
where $r_0$ is the size parameter, while $a$, $b$, and $c$ are the deformation
parameters fixed to $r_0=1$, $a=0.1$, $b=0.01$, and $c=0.012$, respectively \cite{shinohara2010, shinohara2011chaos}. 
As a result of this deformation, the CCW square periodic ray orbit is replaced by a CCW rectangular one, as shown in Fig.~\ref{fig:CAT4PO}(a). Notably, all incident angles remain at $\pi/4$ even after the deformation. 
An orbit with an incident angle slightly above the critical angle densely fills a specific region as it wanders around the CCW rectangular periodic ray orbit, as shown by the red lines in Fig.~\ref{fig:CAT4PO}(a). 
In the Birkhoff coordinates, this trajectory corresponds to four green closed invariant curves surrounding the periodic points, known as stability islands \cite{tabor1989chaos}, as shown in Fig.~\ref{fig:CAT4PO}(b). 
Precisely speaking, several different invariant curves are depicted, each corresponding to a distinct quasi-periodic orbit. Collectively, these curves form the stability islands.
Since both the CCW rectangular periodic orbit and the surrounding islands are located above the critical line 
$\sin \chi=\sin \chi_c$, they are perfectly confined within the deformed optical billiard via total internal reflection.

\begin{figure}[b]
\centering
\includegraphics[width=\columnwidth]{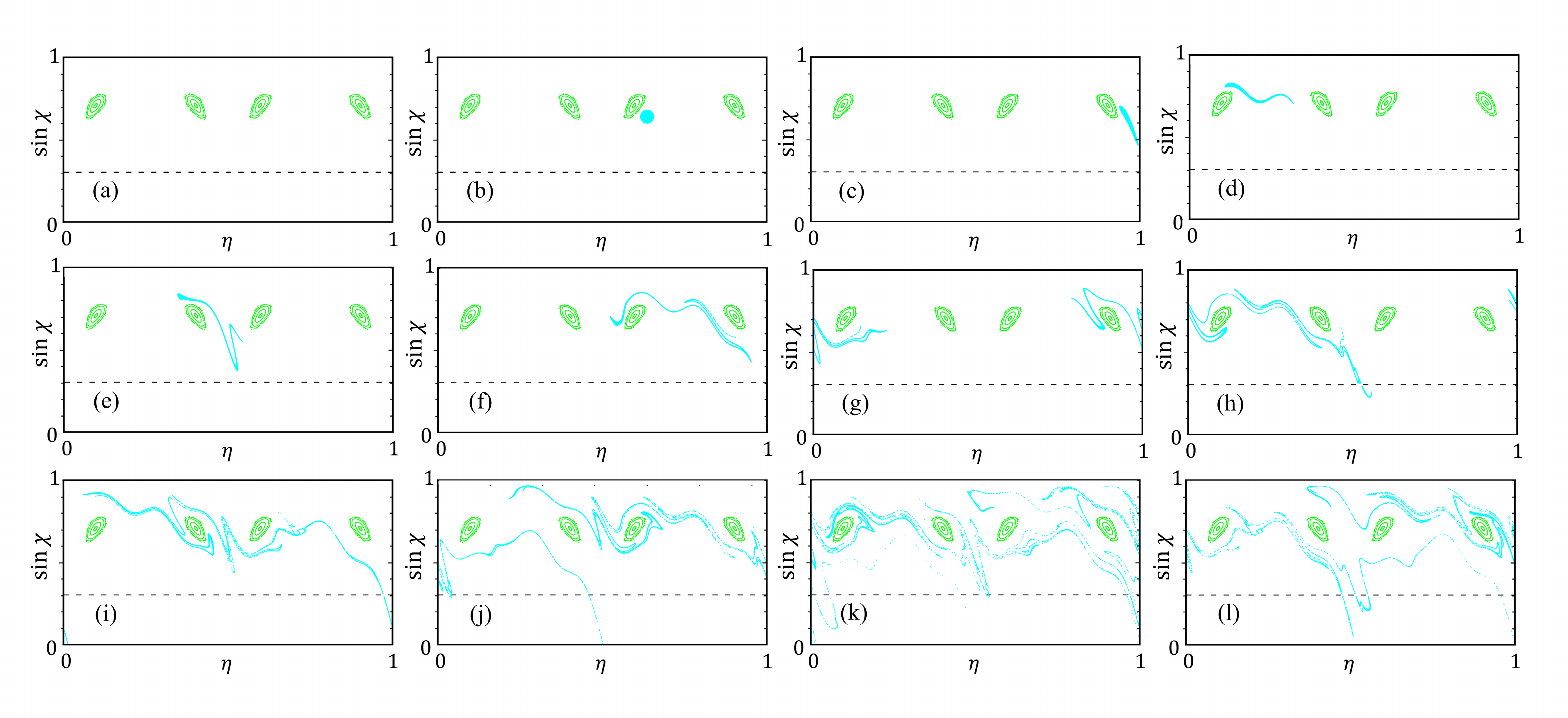}
\caption{
Phase portrait in the Birkhoff coordinates illustrating the time evolution of a chaotic droplet. 
(a) The stability islands of the CCW rectangular periodic orbit.
(b) The initial set of points is represented by a small filled blue circle in the chaotic sea near one of the stability islands. 
(c)--(l) Time evolution of the initial points shown in (b), corresponding to the reflection numbers $1$ to $9$, respectively. 
The droplet is repeatedly stretched and folded as time progresses. The chaotic trajectories diffuse toward the region below the critical line (dashed line) where they eventually impinge on the boundary. The blank regions below the critical line in Fig.~\ref{fig:CAT4PO}, which correspond to the stability islands of the period-2 orbits, act as dynamical barriers that prevent the chaotic orbits from approaching certain $\eta$ ranges, leading to the {\it dynamical eclipsing} phenomenon.
}
\label{fig:CATdynamics}
\end{figure}

Ray orbits starting outside the stability islands wander stochastically in real space, as shown in Fig.~\ref{fig:CAT4PO}(c). In the Birkhoff coordinates, these chaotic trajectories are scattered over a large region known as the chaotic sea shown in Fig.~\ref{fig:CAT4PO}(d). One can observe that these chaotic orbits reach the critical line, implying that they can no longer be perfectly confined within the optical billiard by total internal reflection. The blank regions in Fig.~\ref{fig:CAT4PO}(d) correspond to the stability islands surrounding the periodic orbits with 4 and 2 reflections. It is important to note that chaotic orbits cannot penetrate these stability islands, nor can orbits within the islands escape into the chaotic sea, due to the presence of invariant curves acting as dynamical barriers. Dynamical systems in which stability islands and a chaotic sea coexist in the phase space are referred to as {\it mixed systems} \cite{tabor1989chaos}. In cases where almost all periodic orbits are unstable, the dynamical system is referred to as a {\it fully chaotic system} \cite{tabor1989chaos}.
For almost every initial condition, a typical chaotic trajectory densely samples the accessible phase space in the long-time limit; exceptional marginal orbits form a measure-zero set.
Both mixed systems and fully chaotic systems are categorized as chaotic systems. Accordingly, the deformed optical billiard defined by Eq.~(\ref{eq:cavity}) falls into the category of chaotic optical billiards.

A set of initial points, represented by a small filled blue circle in the chaotic sea near the stability islands of the CCW rectangular periodic orbit in Fig.~\ref{fig:CATdynamics}(b), is repeatedly stretched and folded under time evolution. Whenever some of these points reach the region below the critical line, they impinge on the billiard boundary near $\eta=0$ (or equivalently $\eta=1$) and $\eta=0.5$. This occurs because the two stability islands associated with the period-2 orbits act as obstacles, preventing chaotic trajectories from approaching these regions. This phenomenon is known as {\it dynamical eclipsing} \cite{Noeckel1997}. As a result, chaos-assisted light emission occurs predominantly around $\eta=0$ (and $\eta=1$) and $\eta=0.5$.
Since the emission occurs tangentially to the boundary at these points, bidirectional emission is expected along the upward and downward vertical directions in Fig.~\ref{fig:CAT4PO}(c).

A quantum state localized within a stability island is coupled to the surrounding chaotic sea. This phenomenon is known as {\it dynamical tunneling} \cite{bohigas1993manifestations}, which differs from conventional tunneling through a potential barrier. Once the state tunnels into the chaotic sea, it can spread rapidly and reach other stability islands. In this manner, tunneling between disjoint islands is significantly enhanced by the presence of the chaotic sea, a process referred to as {\it chaos-assisted tunneling} \cite{bohigas1993manifestations}. In the case of an optical billiard, a ray leaking from a stability island into the chaotic sea eventually reaches the critical line and is emitted to the outside. This phenomenon is called {\it chaos-assisted light emission} \cite{shinohara2010, shinohara2011chaos}.


\subsection{Chaotic billiard lasers and resonances in a mixed-phase-space optical billiard}

The integration of a gain medium into a chaotic optical billiard leads to laser action. Such devices are of significant interest from both fundamental and applied perspectives and are known as {\it chaotic billiard lasers} \cite{stone2010chaotic}.
As discussed in Section~\ref{Linear Laser}, under the linear and low-gain approximations, stationary lasing modes can be well approximated by the resonant eigenfunctions of the two-dimensional Helmholtz equation
\begin{equation}
\left(
\nabla^2_{xy} + n^2 k^2  
\right) \varphi 
= 0,
\label{Helmholtz2}
\end{equation}
where $\varphi$ represents the complex amplitude of the $z$-component of the electric field $E_s$ and magnetic field $H_s$ for TM and TE mode, respectively. The boundary conditions at the billiard boundary are given by the continuity of the fields and their derivatives,
\begin{equation}
E_{s,\text{in}}=E_{s,\text{out}}, ~\text{and}~~
\frac{\partial E_{s,\text{in}}}{\partial n}=\frac{\partial E_{s,\text{out}}}{\partial n}, ~~~~~~
\text{for TM},
\label{bdTM}
\end{equation}
and 
\begin{equation}
H_{s,\text{in}}=H_{s,\text{out}}, ~\text{and}~~
\frac{1}{n^2_\text{in}}\frac{\partial H_{s,\text{in}}}{\partial n}=\frac{1}{n^2_\text{out}}\frac{\partial H_{s,\text{out}}}{\partial n}, ~~~~~~
\text{for TE},
\label{bdTE}
\end{equation}
where $\partial/\partial n$ denotes the normal derivative at the billiard boundary. 
At infinity, the Sommerfeld radiation condition is imposed to ensure outgoing waves.
$k$ is the wave number and $n$ is the refractive index, taking the values $n=n_{\text{in}}$ and $n_{\text{out}}$ inside and outside the chaotic billiard laser, respectively. 
In this regime, the spatial distribution of the lasing mode is primarily determined by the passive cavity resonance, allowing us to neglect the nonlinear effects of the gain medium.
The real part of $k$ corresponds to the resonant frequency, while the imaginary part is related to the decay rate (or $Q$-factor) of the mode.

\begin{figure}[b]
\centering
\includegraphics[width=0.6\columnwidth]{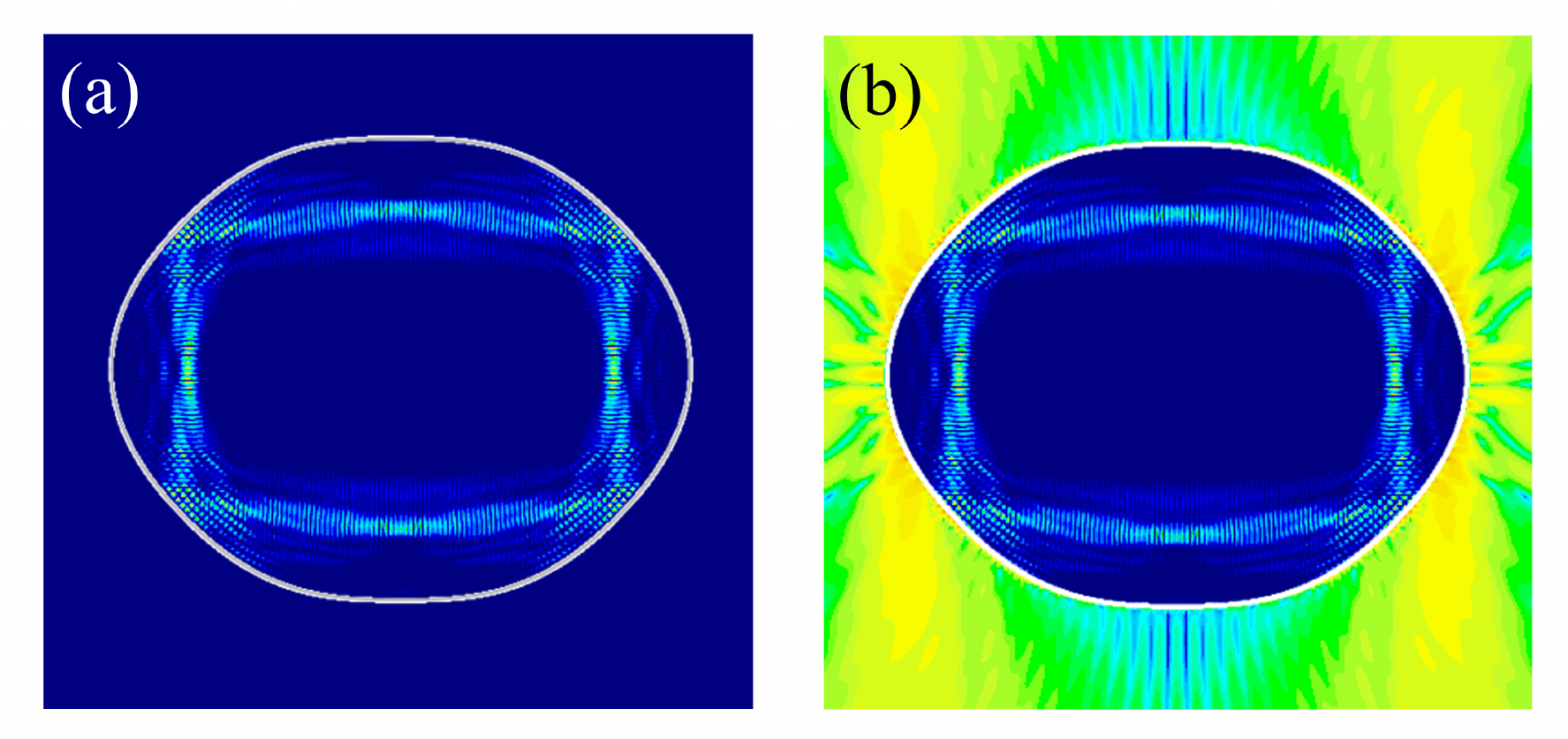}
\caption{
Intensity pattern of a resonance wavefunction localized on the rectangular periodic orbit.
(a) Intensity distribution $|\varphi|^2$ in linear scale for a high-$Q$ TE mode ($k=49.94-i\,0.00012$). 
The mode is strongly localized along the rectangular orbit. (b) The same mode, where the intensity distribution outside the billiard is plotted on a logarithmic scale to highlight the far-field emission pattern, which remains invisible in the linear plot. 
}
\label{fig:CATwt}
\end{figure}

The lasing threshold is closely related to whether the underlying ray orbits of a resonance mode satisfy the total internal reflection condition at the boundary. 
Therefore, the lasing criterion is determined by the critical angle condition for the ray orbits associated with the spatial localization of the resonance mode.
Specifically, modes localized on ray orbits with an incident angle $\chi$ larger than the critical angle $\chi_c$ 
exhibit high $Q$-factors and are thus likely to reach the lasing threshold. 
It is crucial to note that this criterion is applicable only to those chaotic billiard lasers whose resonance wavefunctions exhibit strong localization due to the existence of stability islands, dynamical eclipsing, and other phase-space structures in the ray dynamics. 
Since such ray-wave correspondence becomes invalid for fully chaotic billiard lasers, a nonlinear description incorporating the effects of the active gain medium is essential to accurately describe their lasing characteristics.

Fig.~\ref{fig:CATwt}(a) shows the spatial intensity distribution $|\varphi|^2$ of a high-$Q$ TE resonance mode with a complex wave number $k=49.94-i\,0.00012$, obtained numerically using the boundary element method \cite{wiersig2003boundary}. 
The refractive index inside the billiard is $n_{\text{in}}=3.3$—a typical value of the effective refractive index for semiconductor lasers—while the surrounding air is taken as $n_{\text{out}}=1$. 
Strong localization is clearly observed along the rectangular periodic ray orbit. In this linear-scale plot, the intensity outside the optical billiard appears negligible due to the significantly higher intensity of the localized rectangular orbit. However, the detailed emission pattern is revealed in Fig.~\ref{fig:CATwt}(b), where the intensity outside the billiard is plotted on a logarithmic scale.
One can see bidirectional light emission along the upward and downward vertical directions as ray dynamical analysis predicted.

\subsection{Phase-space representation of resonance modes and chaos-assisted light emission}

\begin{figure}[t]
\centering
\includegraphics[width=\columnwidth]{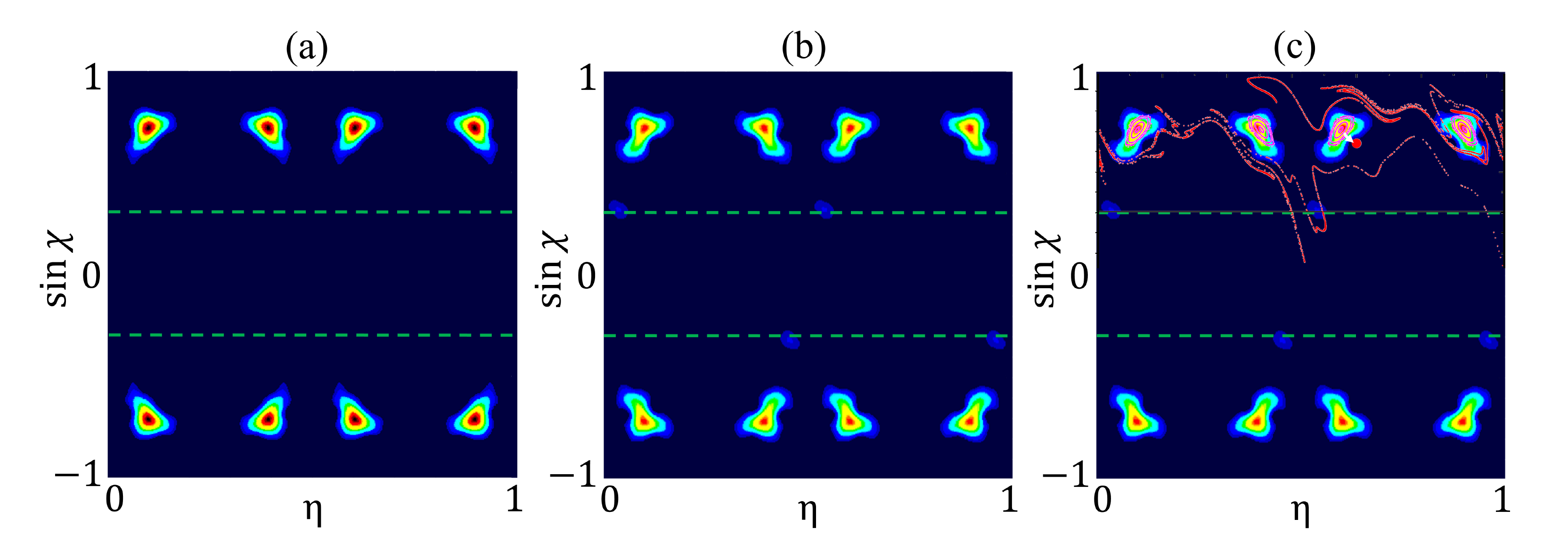}
\caption{
Husimi representation and chaos-assisted light emission. 
(a) Husimi representation of a Dirichlet mode.
The Husimi distribution is calculated for a mode with a wave number close to the resonance mode. The intensity is perfectly localized on the stability islands of the rectangular periodic orbit. 
(b) Husimi representation of the resonance mode.
The Husimi distribution corresponding to the mode in Fig. \ref{fig:CATwt}. In addition to the localization on the rectangular stability islands, intensity is observed near the critical lines around $\eta=0$ (equivalently $\eta=1$) 
 and $\eta=0.5$. This intensity distribution results from the chaotic transport of the tunneled components and the subsequent emission facilitated by the Brewster angle.
(c) Ray-wave correspondence of chaos-assisted light emission. 
The high-intensity regions of the resonance mode near the upper critical line ($\eta \approx 0, 0.5$) are overlapped by the ray distribution, which is shaped by stretching and folding dynamics as well as dynamical eclipsing from the period-2 islands.
}
\label{fig:CATHusimi}
\end{figure}

Fig.~\ref{fig:CATHusimi}(a) shows the Husimi representation of a mode obtained by solving the Helmholtz equation with the Dirichlet boundary condition \cite{crespi1993quantum}. The wave number is chosen to be very close to that of the resonance mode discussed above. This mode is strongly localized within the stability islands of the rectangular periodic ray orbit, and no intensity distribution is observed in the chaotic sea. However, chaos-assisted tunneling occurs between these disjoint stability islands, which lifts the degeneracy and induces a small difference in the wave numbers between modes of different parities.

When the boundary condition is switched to the continuity condition at the billiard boundary and the outgoing wave condition at infinity, the confined light can be emitted from the optical billiard. Fig.~\ref{fig:CATHusimi}(b) shows the Husimi representation of the resonance mode previously shown in Fig.~\ref{fig:CATwt} \cite{hentschel2003husimi}. High-intensity regions appear around $\eta=0$ (equivalently $\eta=1$) and $\eta=0.5$ near the critical line. The portion of the wavefunction leaking from the stability islands into the chaotic sea via dynamical tunneling is transported to these regions by stretching and folding dynamics, combined with dynamical eclipsing caused by the period-2 stability islands, as shown in Fig.~\ref{fig:CATHusimi}(c). A significant portion of the light is then emitted, facilitated by the Brewster angle condition, which results in the observed intensity localization at these specific phase-space locations.

The detail of the light emission can be more precisely analyzed by the flux distribution with Gaussian smoothing; this approach was originally conceived by Prof.~Susumu Shinohara (Komatsu University) \cite{shinohara2007signature}.    
Here, we review the derivation of the flux distribution. According to Poynting’s theorem, the rate of decay of the electromagnetic energy density is expressed by the divergence of the Poynting vector:
\begin{equation}
-\frac{\partial}{\partial t} \left[ \frac{1}{8\pi} (\varepsilon |\bm{E}|^2 + \mu |\bm{H}|^2) \right] = \nabla \cdot \left[ \frac{c}{4\pi} (\bm{E} \times \bm{H}) \right].
\end{equation}
By applying Gauss's divergence theorem, the rate of decay of the total energy 
 within the chaotic billiard laser is expressed as
\begin{equation}
-\frac{dU}{dt}  = \oint_{\partial D} S ds,
\end{equation}
where the total energy $U$ and the energy flux $S$ are defined as 
\begin{equation}
U\equiv \int_{D} \frac{1}{8\pi} (\varepsilon |\bm{E}|^2 + \mu |\bm{H}|^2) dA,
\end{equation}
and 
\begin{equation}
S \equiv  \frac{c}{4\pi} (\bm{E} \times \bm{H})  \cdot \bm{n},
\end{equation}
where $D$ denotes the billiard region and $\bm{n} = ( n_x, n_y,0)$ is the outer normal vector of the billiard boundary. 
For an optical billiard in the TE mode, the electromagnetic fields are given by $\bm{E} = (E_x, E_y, 0)$ and $\bm{H} = (0, 0, H_z)$. Accordingly, $S$ is reduced to 
\begin{equation}
S =  \frac{c}{4\pi} (E_y n_x - E_x n_y)H_z .
\end{equation}

We consider an electromagnetic field oscillating with a complex frequency $\omega = \omega_r + i\omega_i$:
\begin{equation}
H_z=\psi (x,y) e^{-i\omega t}+\psi^* (x,y) e^{i\omega^* t}.
\end{equation}
From Maxwell's equations, we obtain
\begin{equation}
E_x=i\frac{c}{\varepsilon\omega}\frac{\partial \psi}{\partial y} e^{-i\omega t}
-i\frac{c}{\varepsilon\omega^*}\frac{\partial \psi^*}{\partial y} e^{i\omega^* t}  ,
\end{equation}  
and 
\begin{equation}
E_y=-i\frac{c}{\varepsilon\omega}\frac{\partial \psi}{\partial x} e^{-i\omega t}
+i\frac{c}{\varepsilon\omega^*}\frac{\partial \psi^*}{\partial x} e^{i\omega^* t}  .
\end{equation}
Consequently, we arrive at  
\begin{equation}
S=\frac{c^2}{2\pi\varepsilon}\operatorname{Im}\left(\frac{1}{\omega}\psi^\ast\frac{\partial \psi}{\partial n}+\frac{1}{\omega}\psi\frac{\partial \psi}{\partial n}e^{-i{2\omega}_rt}\right)e^{{2\omega}_it}  .
\end{equation}
The rapidly oscillating component with frequency $2\omega$ is averaged out by introducing the time-averaged flux $\bar{S}$ 
\begin{equation}
\bar{S}(\eta,t) \equiv \frac{\omega_r}{2\pi}\int_{t}^{t+\frac{2\pi}{\omega_r}}{Sdt^\prime}\cong 
\frac{c^2}{2\pi\varepsilon}\operatorname{Im}\left(\frac{1}{\omega}\psi^\ast\left(\eta\right)\frac{\partial \psi\left(\eta\right)}{\partial n}\right)e^{{2\omega}_it},
\end{equation}
and by applying the same averaging to the total energy we obtain 
\begin{equation} 
\omega_i\cong -\frac{c^2}{4\pi n^2 \omega_r}
\oint_{\partial D} d\eta
\operatorname{Im}\left(\psi^\ast\left(\eta\right)\frac{\partial \psi\left(\eta\right)}{\partial n}\right),
\end{equation}
under the assumption of the normalization of the stationary total energy $U_s\cong e^{-2\omega_i t}\int_t^{t+\frac{2\pi}{\omega_r}}dt U$.

The rapid spatial oscillations are further smoothed out by introducing a Gaussian filter:
\begin{equation}
\bar{\bar{S}}(\eta,t)\equiv\oint_{\partial {D}}{d\eta^\prime\ \bar{S}(\eta^\prime,t)K_L(\eta,\eta^\prime,\sigma\mathrm{)}},
\end{equation}
where $K_L$ is the Gaussian kernel defined on a one-dimensional torus of length $L$,
\begin{equation}
K_L\left(\eta,\eta^\prime,\sigma\right)\equiv\sum_{m=-\infty}^{\infty}{\frac{1}{\sqrt{2\pi}\sigma}
\exp{\left[-\frac{\left(\eta-\eta^\prime+mL\right)^2}{2\sigma^2}\right]}}.     
\end{equation}
Note that while $L=1$ for the Birkhoff coordinates, we keep the explicit dependence on $L$ for generality and clarity.
Applying the identity,
\begin{equation}
\frac{1}{2\pi}\int_{-\infty}^{\infty}dp G_L^\ast\left(\eta,\eta^{\prime\prime},p,\sqrt2\sigma\right)G_L\left(\eta,\eta^\prime,p,\sqrt2\sigma\right)  
=K_L\left(\eta,\eta^{\prime\prime},\sigma\right)\delta\left(\eta^{\prime\prime}-\eta^\prime\right), 
\end{equation}
to the flux $\bar{\bar{S}}$,
\begin{multline}
\bar{\bar{S}}(\eta,t)=\frac{c^2}{2\pi\varepsilon}e^{{2\omega}_it}\oint_{\partial D}\ d\eta^\prime   
\operatorname{Im}\left(\oint_{\partial D}\ d\eta^{\prime\prime}\delta\left(\eta^{\prime\prime}-\eta^\prime\right)
\frac{1}{\omega}\psi^\ast\left(\eta^\prime\right)\frac{\partial \psi\left(\eta^{\prime\prime}\right)} 
{\partial n}K_L(\eta,\eta^{\prime\prime},\sigma)\right) \hfill \\ 
\ \ \ \ \ \ \ \ \ \ \ \ \ \ \ =\int_{-\infty}^{\infty}dp\frac{c^2}{4\pi^2\varepsilon}e^{2\omega it}
\operatorname{Im}\biggl(\frac{1}{\omega}\oint_{\partial D}{G_L\left(\eta,\eta^\prime,p,\sqrt2\sigma\right)\psi^\ast\left(\eta^\prime\right)d\eta^\prime}  
\oint_{\partial D}{d\eta^{\prime\prime}
G_L^\ast\left(\eta,\eta^{\prime\prime},p,\sqrt2\sigma\right)}\frac{\partial \psi\left(\eta^{\prime\prime}\right)}{\partial n}
\biggr),  \hfill
\end{multline}
where $G_L$ is the minimum uncertainty wave packet on a one-dimensional torus:
\begin{equation}
G_{L\ }\left(\eta,\eta^\prime,p,\sqrt2\sigma\right)= 
\left(\frac{1}{2\pi\sigma^2}\right)^\frac{1}{4}\sum_{m=-\infty}^{\infty}
\exp{\left[-\frac{\left(\eta-\eta^\prime+mL\right)^2}{4\sigma^2}+ip\left(\eta-\eta^\prime+mL\right)\right]} .
\end{equation}
Finally, we obtain the flux distribution $P(\eta,\sin\chi)$ for a TE resonance mode $\psi$,
\begin{equation}
P(\eta,\sin\chi)\equiv
\frac{c^2}{4\pi^2\varepsilon}\operatorname{Im}\left(\frac{1}{\omega}\Psi_\psi^\ast{\left(\eta,\sin\chi\right)\Psi}_{\partial_n \psi}\left(\eta,\sin\chi\right)\right),
\end{equation}
where 
\begin{equation}
\Psi_\psi^\ast=\oint_{\partial D}{G_L\left(\eta,\eta^\prime,\sin\chi,\sqrt2\sigma\right)\psi^\ast\left(\eta^\prime\right)d\eta^\prime},
\end{equation}
and
\begin{equation}
\Psi_{\partial_n \psi}=\oint_{\partial D}{d\eta^{\prime\prime} G_L^\ast
\left(\eta,\eta^{\prime\prime},\sin \chi,\sqrt2\sigma\right)}\frac{\partial \psi\left(\eta^{\prime\prime}\right)}{\partial n}.
\end{equation}
The flux distribution $P(\eta,\sin\chi)$ for a TM-mode resonance can be obtained in an analogous manner \cite{shinohara2007signature}.

\begin{figure} [t]
\centering
\includegraphics[width=0.4\columnwidth]{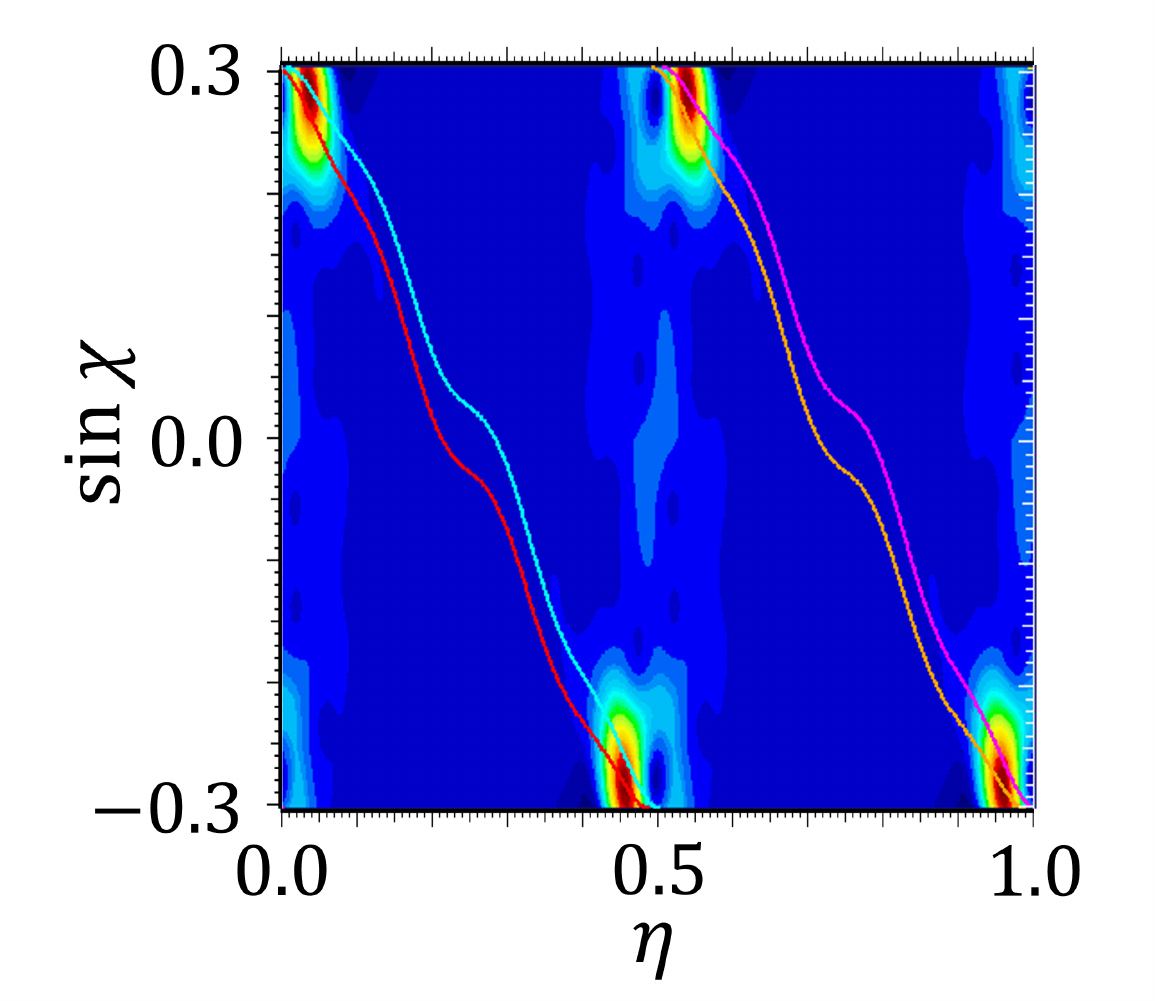}
\caption{
Flux distribution with Gaussian smoothing, as conceived by Prof. Susumu Shinohara. 
The distribution is shown as a function of the boundary position $\eta$. 
The upper and lower boundaries of the plot correspond to the critical lines, 
representing the condition where the emission is tangential to the billiard edge 
(emission angle $\pm\pi/2$). The strong localizations observed just inside these critical lines 
(e.g., $\eta\approx 0.03, 0.47,0.53, 0.97 $) indicate the emergence of highly directional light emission around $\pm 90^\circ$.
The four curves represent the relationship between the boundary position $\eta$ and 
the incident angle $\chi$ for which the emission angles are exactly $\pm(90\pm1.6)^\circ$.
}
\label{fig:CATFlux}
\end{figure}

This distribution provides precise information regarding the near-field of light emission from a chaotic billiard laser.
Figure~\ref{fig:CATFlux} displays the flux distribution for the resonance mode shown in Fig.~\ref{fig:CATwt}.
The strong localizations near the critical lines, specifically around $\eta=0.03$ and $0.97$ 
(as well as $0.47$ and $0.53$), imply that pairs of closely spaced bright spots are expected to be observed near the right (and left) edges of the chaotic billiard laser.

These critical lines correspond to the incident angles at which the emission angle is $\pm\pi/2$, 
representing emission tangential to the billiard boundary.
The localization at $\eta=0.03$, situated just below the upper critical line, indicates intense directional emission 
near 90$^\circ$, while the localization at $\eta=0.97$, just above the lower critical line, suggests emission toward $-90^\circ$. 
Similarly, the concentrations at $0.47$ and $\eta=0.53$ correspond to directional emissions at $90^\circ$ and $-90^\circ$, respectively. Due to the spatial separation between the two light-emitting points for the directional emissions at $\pm90^\circ$, significant interference patterns emerge in the far field patterns.

Furthermore, the four curves in Fig.~\ref{fig:CATFlux} represent the relationship between the boundary position $\eta$ and 
the incident angle $\chi$ for which the emission angles are exactly $\pm(90\pm1.6)^\circ$. 
Although these curves pass through the maximum value of the flux distribution, the values in the vicinity of the maximum point are comparable to the maximum itself. Accordingly, these slight differences between $\pm90^\circ$ and $\pm(90\pm1.6)^\circ$ are obscured by the rapid oscillations arising from the interference between the two spatially separated light-emitting points. 
However, the high-frequency interference fringes are suppressed when the far-field pattern is integrated over the detector’s active area in the real experiment.
Therefore, the significant overlap between these curves and the localized flux distribution confirms that the smoothed far-field pattern will exhibit prominent peaks at $\pm90^\circ$ with sub-peaks at $\pm(90\pm1.6)^\circ$.

\section{Experiments on semiconductor chaotic billiard lasers}
\label{Semion}

\begin{figure}[b]
\centering
\includegraphics[width=0.6\columnwidth]{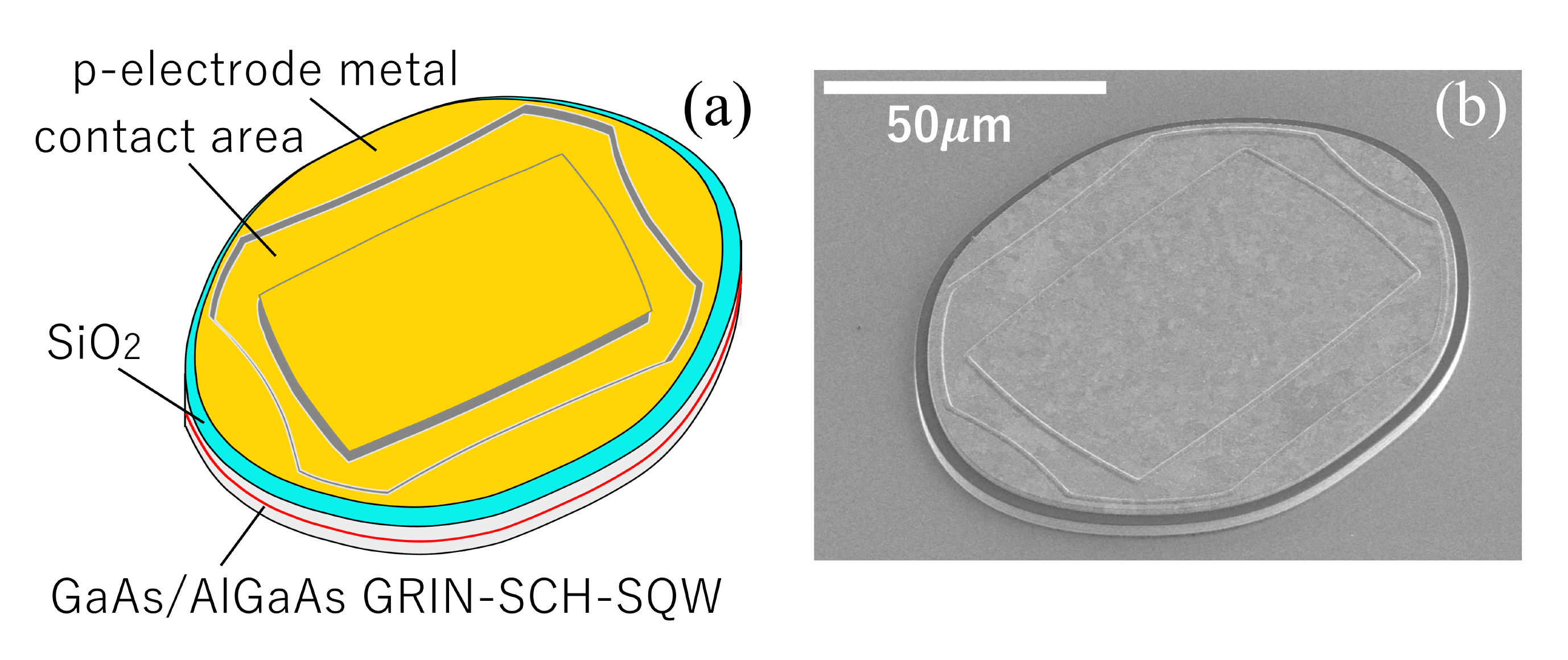}
\caption{
Device structure and fabrication results. 
(a) Schematic of the upper epitaxial layers and the electrode contact window patterned along the rectangular orbit, 
which features inwardly recessed segments. For clarity, the $n$-GaAs substrate and the underlying $n$-electrode metal are not shown. 
(b) Scanning electron microscope (SEM) image of a fabricated laser diode with $M=0.95$. 
The recessed structure of the electrode, corresponding to the design in (a), is clearly visible.
Panel (b) is adapted with permission from \cite{shinohara2010}. Copyright 2010 by the American Physical Society.
}
\label{fig:CATSch}
\end{figure}

\subsection{Fabrication of semiconductor chaotic billiard lasers}
\label{Fabrication}

The laser diodes were fabricated using a graded-index separate-confinement-heterostructure (GRIN-SCH) single-quantum-well (SQW) GaAs/AlGaAs structure, which was grown by metal-organic chemical vapor deposition (MOCVD) \cite{shinohara2010, shinohara2011chaos}. This GRIN-SCH configuration, featuring a single quantum well at its center, provides efficient optical and carrier confinement for the lasing mode.
The devices were fabricated from MOCVD-grown epitaxial layers, consisting of a 1.5-$\mu$m $n$-Al$_{0.5}$Ga$_{0.5}$As lower cladding layer, a 0.2-$\mu$m $n$-Al$_x$Ga$_{1-x}$As ($x=0.5-0.2$) graded layer, and a 10-nm GaAs single quantum well.
This was followed by a 0.2-$\mu$m $p$-Al$_x$Ga$_{1-x}$As($x=0.2-0.5$) graded layer, a 1.5-$\mu$m 
$p$-Al$_{0.5}$Ga$_{0.5}$As upper cladding layer, and a 0.2-$\mu$m $p$-GaAs contact layer. 
Additionally, a 400-nm-thick SiO$_2$ layer was deposited.
The lasing wavelength is approximately 850 nm.
The average radius $r_0$ of the device is fixed as 50 $\mu$m. 
The effective refractive index is $n_{\text{in}}=3.3$.

To selectively inject current into the rectangular periodic ray orbit modes, the electrode contact window was patterned along the rectangular orbit, which is characterized by inwardly recessed segments. The specific geometry of this contact window is illustrated in Fig. \ref{fig:CATSch}(a). 
A scanning electron microscope (SEM) image of a fabricated laser diode with $M=0.95$ 
is shown in Fig. \ref{fig:CATSch}(b), where the recessed structure of the electrode can be clearly observed.
For current injection, the top of the cavity was coated with an electrode metal, covering the entire surface except for a small margin.
The precise shape of the electrode metal is defined by $r<M r(\phi)$, where $M (\leq 1)$ is the margin parameter. 
Samples with $M=0.8, 0.85$ and $0.95$ were fabricated.
The SiO$_2$ layer acts as an insulation barrier between the semiconductor layers and the $p$-electrode metal. 
This ensures that the electrode only makes contact with the GaAs layer through the patterned window, 
thereby enabling selective current injection. 
Such a configuration allows for the preferential excitation of rectangular periodic ray orbit modes while effectively suppressing WG-type modes.

\begin{figure}[b]
\centering
\includegraphics[width=0.6\columnwidth]{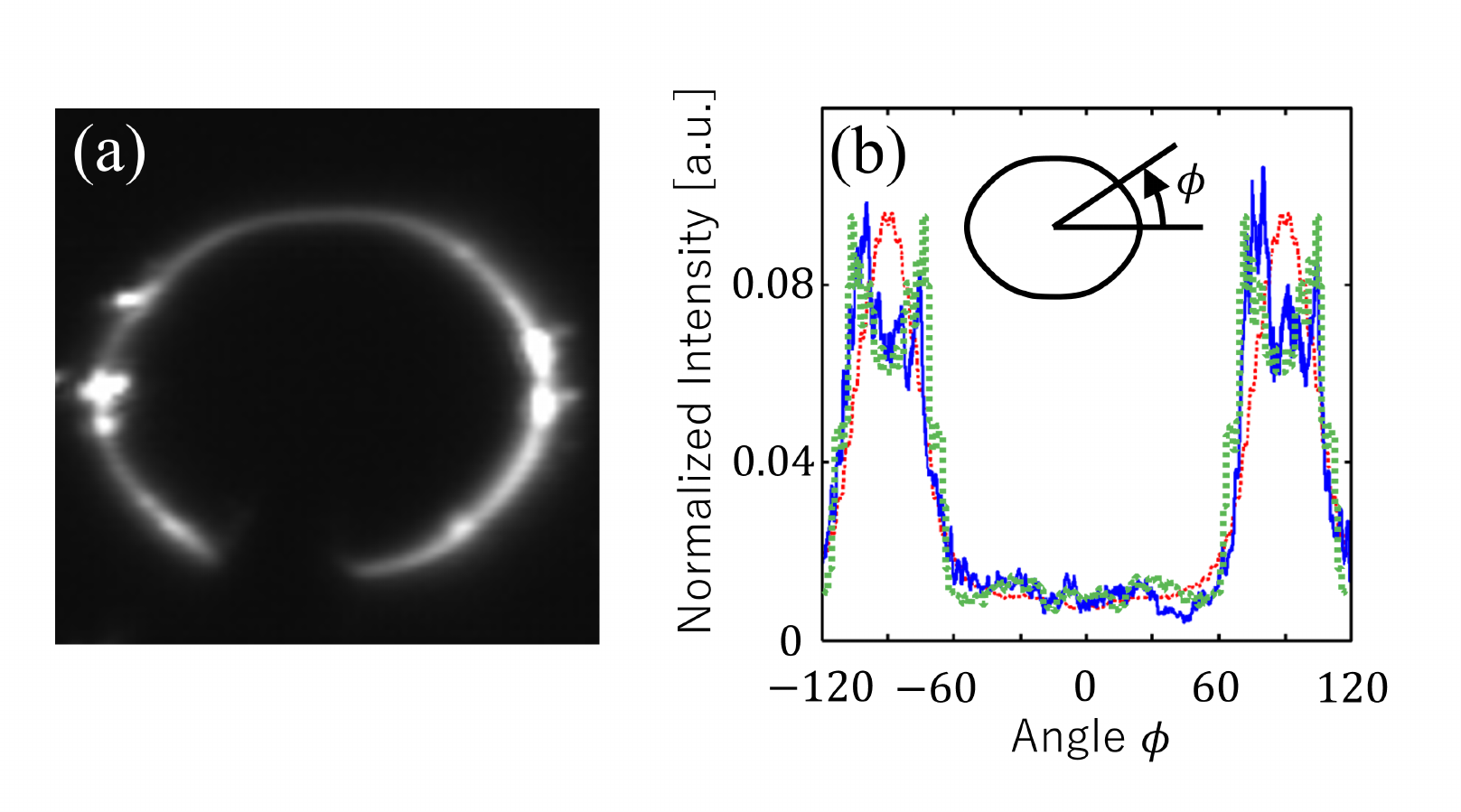}
\caption{
Near-field and far-field emission characteristics. (a) Near-field pattern (NFP) of the cavity during lasing operation, captured by an IR-CCD camera. The two distinct emission points on each side of the cavity are consistent with the theoretical predictions based on the phase-space representation and chaos-assisted light emission. Note that the dark region on the cavity boundary is caused by the probe used for current injection.  (b) Far-field emission patterns (FFP): experimental data (solid blue curve) at 500 mA vs. ray-calculated (dashed green curve) and wave-calculated (dotted red curve) results. 
In the wave calculation, rapid interference-induced oscillations are artificially smoothed out.
Figures are adapted with permission from \cite{shinohara2010} and \cite{shinohara2011chaos}. Copyright 2010 and 2011 by the American Physical Society.
}
\label{fig:NFPFFP}
\end{figure}

\subsection{Experimental observation of chaos-assisted light emission}
\label{Experiment}

We tested the laser diodes at 25$^{\circ}$C using a pulsed current (500-ns width at 1-kHz repetition rate).
Figure \ref{fig:NFPFFP}(b) shows the measured far-field pattern (FFP) for the diode with $M=0.95$ at an injection current of 500 mA, alongside both ray- and wave-calculated FFPs corresponding to the mode shown in Fig. \ref{fig:CATwt}. 
The data are normalized to unit area.
Prior to the normalization of the experimental data, the uniform background contribution arising from spontaneous emission was subtracted, ensuring that the emission level near $\phi=0$ matches the wave-calculated data.

As shown in Fig.~\ref{fig:NFPFFP}(b), the experimental results are in excellent agreement with the numerical calculations. 
The far-field patterns for samples with $M=0.8$ and $0.85$ exhibit similar overall profiles to those for $M=0.95$, 
although slight differences appear in the sub-structures of the two main peaks.
At an injection current of 500 mA, optical spectra confirmed that lasing occurs in multiple longitudinal modes associated with the rectangular periodic ray orbit. The sub-structures observed in the experimental main peaks are attributed to this multi-mode operation, resulting in sample-specific variations.
In the wave-calculated FFP shown in Fig.~\ref{fig:NFPFFP}(b), the rapid oscillations arising from interference effects have been artificially smoothed out to facilitate a clearer comparison.
Despite this smoothing, the ray-calculated FFP appears to provide a better fit to the experimental data than the wave-calculated one. The underlying physical mechanism—specifically, how multi-mode lasing effectively reproduces the ray-calculated FFP through an averaging process—remains an open question and is a subject for future investigation \cite{shinohara2008light, choi2008dependence, harayama2015ray, ketzmerick2022chaotic}.

To identify the specific locations of light emission, we performed near-field measurements of the cavity during lasing operation using an infrared (IR) CCD camera. The result for the sample with $M=0.95$  is shown in Fig. \ref{fig:NFPFFP}(a), where scattered light is observed at the cavity boundaries. Consistent with the theoretical predictions discussed in the previous section on the phase-space representation of resonance modes and chaos-assisted light emission, we can clearly observe that light is emitted from two distinct points on each side of the cavity.

\begin{figure}[t]
\centering
\includegraphics[width=\columnwidth]{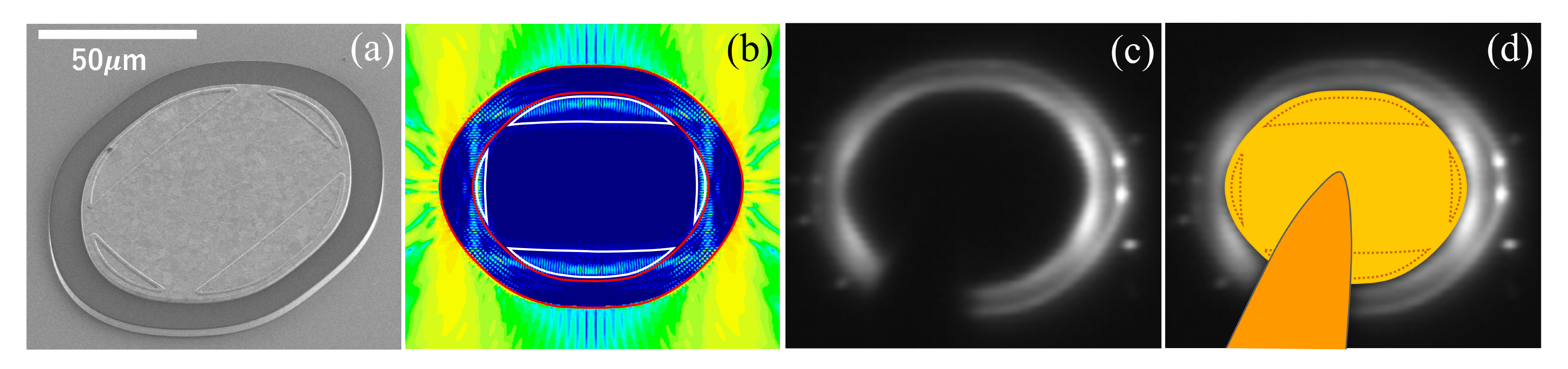}
\caption{
Experimental observation of modal patterns and chaos-assisted light emission. 
(a) Scanning electron microscope (SEM) image of the fabricated laser diode with $M=0.8$. 
(b) Superimposed image of the calculated intensity pattern of the rectangular periodic ray orbit, the contact window (white curves), and the SiO$_2$ margin area (outside the red curve). 
(c) Near-field pattern (NFP) captured by an IR-CCD camera during lasing operation. The internal modal pattern observed through the SiO$_2$ margin closely matches the rectangular orbit in (b). 
(d) Schematic illustration showing the configuration of the $p$-electrode metal, the contact window, and the probe used for current injection.
Figures are adapted with permission from \cite{shinohara2010} and \cite{shinohara2011chaos}. Copyright 2010 and 2011 by the American Physical Society.
}
\label{fig:CATLast}
\end{figure}

To further investigate the modal patterns of the lasing modes, the SiO$_2$ margin was extended to $M=0.8$, allowing the internal intensity distribution to be observed through the transparent insulation layer. Lasing was performed under the same conditions as in the previous measurements.
Figure \ref{fig:CATLast}(a) shows the SEM image of the device, 
while Fig.~\ref{fig:CATLast}(b) illustrates the calculated intensity pattern of the rectangular periodic ray orbit, with the contact window (white curves) and the SiO$_2$ margin (red curve) superimposed. 
The corresponding near-field pattern (NFP) captured by the IR-CCD camera is shown in Fig.~\ref{fig:CATLast}(c). 
For clarity, the configuration of the $p$-electrode, the probe for current injection, 
and the contact window is depicted in Fig.~\ref{fig:CATLast}(d).
In addition to the two bright emission spots at the cavity edges, a distinct modal pattern is observed within the SiO$_2$ margin, which closely coincides with the rectangular periodic ray orbit mode. These observations lead us to conclude that current is selectively injected into the rectangular orbit area, where dynamical tunneling occurs from the stability islands into the surrounding chaotic sea. Consequently, the laser light is emitted from the cavity, providing direct experimental evidence of chaos-assisted light emission in semiconductor chaotic billiard lasers.

\section{Fully chaotic billiard lasers}
\label{FCBL}

\subsection{Fully chaotic billiards}
\label{Full chaos}

In the previous sections, we noted that the critical angle for total internal reflection plays a pivotal role in determining the lasing characteristics of partially chaotic billiard lasers. In such systems, light confinement is achieved through the existence of stability islands above the critical line, while emission is mediated by the stretching and folding dynamics of phase space and dynamical eclipsing.

In contrast, in a fully chaotic billiard, all periodic orbits are unstable. Almost all ray trajectories originating above the critical line eventually scatter to angles below the critical threshold after several reflections, wandering randomly across the billiard table. Consequently, the conventional critical angle condition becomes ineffective as a criterion for lasing in fully chaotic systems.

\begin{figure}[t]
\centering
\includegraphics[width=\columnwidth]{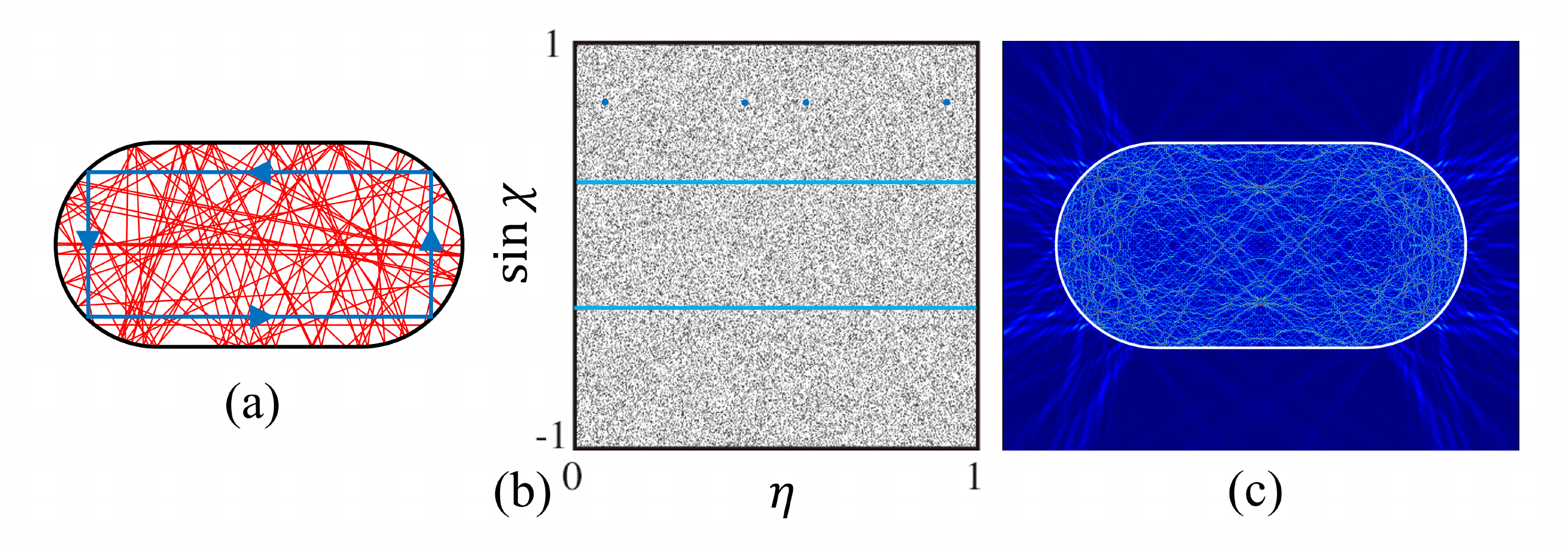}
\caption{
Ray dynamics and a resonance in a fully chaotic stadium billiard. (a) Trajectories of light rays: the blue lines indicate the unstable CCW rectangular periodic orbit with an incident angle of $\pi/4$, while the red lines represent a typical chaotic trajectory that rapidly departs from the periodic path. (b) Phase-space representation (Birkhoff coordinates). The blue horizontal lines denote the critical lines for total internal reflection. In this fully chaotic system, a single ray trajectory (represented by black dots) eventually covers the entire phase space, regardless of the critical angle condition. 
The unstable CCW rectangular periodic orbit is marked by four blue dots.
(c) Calculated resonance wave function in a fully chaotic stadium optical billiard. The spatial distribution of the electromagnetic field intensity is shown. Unlike partially chaotic billiards, the wave function exhibits no clear localization along specific periodic orbits and spreads over the entire cavity, illustrating the breakdown of the conventional critical angle confinement.
}
\label{fig:Stad4POChaos}
\end{figure}

The stadium billiard serves as a canonical example of such fully chaotic dynamics. The counter-clockwise (CCW) rectangular periodic orbit with an incident angle of $\pi/4$ as in circular or deformed billiards is indicated by the four blue lines in Fig.~\ref{fig:Stad4POChaos}(a). The corresponding orbit represented in Fig.~\ref{fig:Stad4POChaos}(b) is located above the upper critical line (The effective refractive index is $n_{\text{in}}=3.3$.). However, this orbit is inherently unstable; any trajectory with a slight initial deviation rapidly departs from the periodic path and wanders stochastically, as illustrated by the red lines in Fig.~\ref{fig:Stad4POChaos}(a) and the phase-space distribution in Fig.~\ref{fig:Stad4POChaos}(b).
 
%
Similarly, the resonance wave functions in fully chaotic optical billiards exhibit broad spatial distributions without localization (\ref{fig:Stad4POChaos}(c)), further indicating that the critical angle condition is inapplicable. If this condition were a strictly necessary requirement for lasing, one would conclude that laser action in a fully chaotic stadium billiard is impossible. However, it is crucial to recognize that the critical angle condition is a sufficient, rather than a necessary, condition for ensuring that the modal gain exceeds the total losses.

\begin{figure}[b]
\centering
\includegraphics[width=0.6\columnwidth]{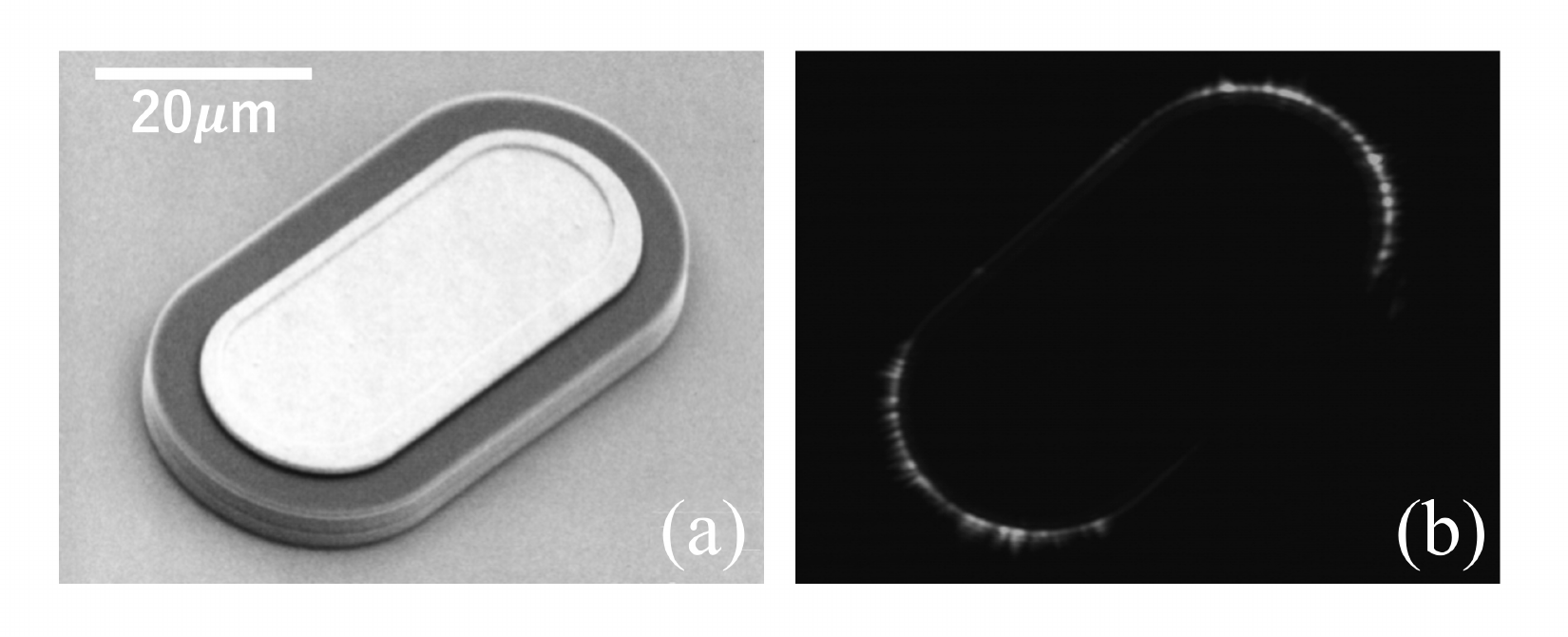}
\caption{
Experimental observation of laser action in a fully chaotic stadium billiard laser. (a) Scanning electron microscope (SEM) image of the fabricated device. The contact window for uniform current injection covers the entire area of the stadium. (b) Near-field image captured by an IR-CCD camera during lasing operation. Despite the chaotic nature of the cavity and the violation of the total internal reflection condition, stable laser action is confirmed by the distinct threshold behavior and spectral characteristics.
Panel (a) is adapted with permission from \cite{Fukushima2004}. \copyright\ 2004 IEEE.
}
\label{fig:StadExp}
\end{figure}

Indeed, we have successfully fabricated a semiconductor fully chaotic stadium billiard laser, as shown in the SEM image in Fig.~\ref{fig:StadExp}(a). The fabrication process and device structure are essentially identical to those of the partially chaotic lasers, though uniform pumping was applied via a contact window covering the entire billiard area (excluding the margin). Laser action was unequivocally confirmed through the observation of a clear current threshold, the linear increase of emission intensity with injection current, spectral analysis, and the far- and near-field emission profiles (Fig.~\ref{fig:StadExp}(b)) \cite{Fukushima2004,  shinohara2008light, choi2008dependence, sunada2013stable, sunada2016signature, you2022universal}.

The observation that lasing occurs despite the violation of the critical angle condition necessitates a distinct theoretical framework for fully chaotic lasers. In the following sections, we present a nonlinear theory that accounts for all orders of electromagnetic field nonlinearity, which is essential to understanding how the gain compensates for the inherent losses in these systems \cite{harayama2011two, harayama2005theory, harayama2003stable}.

\subsection{Maxwell-Bloch equations}



In subsection \ref{MB eqs}, Maxwell equations for optical billiards were derived. By incorporating the internal loss coefficient $\beta$, the wave equations for the TM and TE modes are given by
\begin{equation}
    \left(\nabla_{x y}^{2}-\frac{n^2}{c^2} \frac{{\partial}^2}{\partial t^2} \right)  E_z
    =2\beta \frac{n^2}{c^2}\frac{{\partial} E_z}{\partial t} + \frac{4 \pi}{c^2} \frac{{\partial}^2 P_z}{\partial t^2},
    \label{eq1}
\end{equation}
and 
\begin{equation}
\left(\nabla_{x y}^{2}-\frac{n^2}{c^2} \frac{{\partial}^2}{\partial t^2} \right)H_z=2\beta \frac{n^2}{c^2}\frac{{\partial} H_z}{\partial t} 
-\frac{4\pi }{c}\frac{\partial}{\partial t}\left(\frac{\partial P_y}{\partial x}-\frac{\partial P_x}{\partial y}\right), 
\label{eq2}
\end{equation}
respectively.
The loss coefficient $\beta$ is introduced phenomenologically to represent absorption within the cavity, while it vanishes outside.  The refractive index $n(x, y)$ is defined as $n_{\text{in}}$ inside the cavity and $n_{\text{out}}$ outside. 
In the above, we neglected the $z$-dependence of the electromagnetic field 
by assuming it is confined within a small vertical extent, 
such as a semiconductor quantum well. This allows the field to be approximated by the fundamental eigenstate 
along the $z$-direction. A separation of TM and TE modes, independent of the $z$-coordinate, 
can also be achieved in structures that are uniform along the $z$-axis, such as droplets or jets (see Subsection \ref{thin film}). 


Next, we describe the dynamics of the lasing medium consisting of two-level atoms \cite{sargent1974laser}. We assume that each atom possesses spherical symmetry and has two energy levels that correspond to the states $|s\rangle$ and $|p_{i}\rangle, i \in {x,y,z}$.
For TM-polarized fields, where the electric field possesses only a $z$-component, the interaction Hamiltonian is given by
\begin{equation}
H_I = -e \rr \cdot \EE = -e z E_z = -\hat{d} E_z,
\end{equation}
where $\rr$ and $e$ denote the position and charge of the electron, respectively, and 
$\hat{d}$ represents the $z$-component of the dipole moment operator.
In this case, the relevant atomic states can be described as 
a superposition of the states $|s\rangle$ and $|p_z\rangle$:
\begin{equation}
|\Psi\rangle = c_s |s\rangle + c_z |p_z\rangle.
\end{equation}
This is because the transition matrix elements between the states $|s\rangle$ and $|p_z\rangle$ 
vanish due to the symmetry of the atomic orbitals and the specific $z$-polarization of the electric field.

Accordingly, the total Hamiltonian of the atom-field system is given by
\begin{equation}
H = \hbar \left( \sum_{i=x,y,z} \omega_{+} |p_i\rangle \langle p_i| + \omega_{-} |s\rangle \langle s| \right) - \hat{d} E_z.
\end{equation}
Expanding the dipole operator $\hat{d}$, the Hamiltonian can be expressed as
\begin{equation}
H = \hbar \left( \sum_{i=x,y,z} \omega_{+} |p_i\rangle \langle p_i| + \omega_{-} |s\rangle \langle s| \right) - (d_{sz} |s\rangle \langle p_z| + d_{sz} |p_z\rangle \langle s|) E_z,
\end{equation}
where $d_{sz}$ denotes the transition dipole moment between the two states $|s\rangle$ and $|p_z\rangle$. 
The temporal dynamics of the atomic states $\Psi$ is then governed by the Schrödinger equation:
\begin{equation}
i\hbar \frac{\partial}{\partial t} |\Psi\rangle = H |\Psi\rangle.
\label{SchrodingerEq}
\end{equation}
Substituting the state vector $\Psi=c_s |s\rangle + c_z |p_z\rangle$ into Eq.~(\ref{SchrodingerEq}), 
we obtain the coupled equations for the time evolution of the probability amplitudes $c_s$ and $c_z$ 
\begin{equation}
\dot{c}_s = -i \omega_{-} c_s + i \kappa c_z E_z,
\label{evolution_cs}
\end{equation}
and
\begin{equation}
\dot{c}_z = -i \omega_{+} c_z + i \kappa c_s E_z.
\label{evolution_cz}
\end{equation}
Here, for simplicity, we define the coupling constant as $\kappa\equiv d_{sz}/\hbar$.
By defining the microscopic polarization as $\rho \equiv c_s^*c_z$ and the population inversion as $W \equiv |c_z|^2 -|c_s|^2$, 
we obtain the following equations governing their temporal evolution
\begin{equation}
\frac{\partial \rho}{\partial t} = - i \omega_0 \rho - i \kappa W E_z,
\end{equation}
and
\begin{equation}
\frac{\partial W}{\partial t} = -2 i \kappa E_z (\rho - \rho^*),
\end{equation}
where $\omega_0\equiv \omega_{+}-\omega_{-}$ is the transition frequency. 
To generalize these results to the cases where the electric field $\EE$ points in an arbitrary direction, 
$E_z$ can be replaced by $E\equiv \ee \cdot \EE$, where the unit vector $\ee$ is oriented parallel to the electric field $\EE$.

Relaxation due to interaction with the reservoir is phenomenologically incorporated using two decay constants: 
the transverse relaxation rate $\gamma_\perp$ for microscopic polarization 
and the longitudinal relaxation rate $\gamma_{\parallel}$ for population inversion. 
Additionally, the effect of injection of external energy into the lasing medium is represented by the pumping power $W_{\infty}$.
Finally, we obtain the optical Bloch equations for the atoms in the lasing medium
\begin{equation}
\frac{\partial \rho}{\partial t} = -(i \omega_0 + \gamma_{\perp}) \rho - i \kappa W E,
\end{equation}
and
\begin{equation}
\frac{\partial W}{\partial t} = -2 i \kappa E (\rho - \rho^*) - \gamma_{\parallel}(W - W_{\infty}).
\label{MB_final}
\end{equation}

Microscopic polarization $\pp$ is defined as the expectation value of the dipole operator:
\begin{equation}
\pp \equiv \langle\Psi| e \rr |\Psi\rangle.
\end{equation}
For the $z$-component, we obtain
\begin{equation}
p_z = \langle\Psi| (d_{sz} |s\rangle\langle p_z| + d_{sz}|p_z\rangle\langle s|) |\Psi\rangle
= (c_s^* c_z + c_z^* c_s) \hbar \kappa
= (\rho + \rho^*) \hbar \kappa,
\end{equation}
where we have used the relation $d_{sz}=\hbar\kappa$. 
This result is straightforwardly extended to the general case where the electric field has all three components $(E_x,E_y,E_z)$. Specifically, we obtain
\begin{equation}
\pp
=  \kappa \hbar (\rho + \rho^*)\ee,
\end{equation}
where $\ee$ denotes the unit vector in the direction of the electric field. Consequently, the macroscopic polarization $\PP$ is given by
\begin{equation}
 \PP = N \pp = N \kappa \hbar \left( \rho + \rho^* \right)\e.
\end{equation}
where $N$ is the number density of atoms.


Consequently, we arrive at the complete set of Maxwell-Bloch equations for the TM mode:
\begin{equation}
    \left(\nabla_{x y}^{2}-\frac{n^2}{c^2} \frac{{\partial}^2}{\partial t^2} \right)  E_z
    =2\beta\frac{n^2}{c^2} \frac{{\partial} E_z}{\partial t} + \frac{4 \pi}{c^2} \frac{{\partial}^2 P_z}{\partial t^2},
    \label{Eq1}
\end{equation}
\begin{equation}
P_z = N \hbar \kappa (\rho + \rho^*),
\label{P-rho-relation}
\end{equation}
\begin{equation}
\frac{\partial \rho}{\partial t} = -(i \omega_0 + \gamma_\perp) \rho - i \kappa W E_z,
\label{MB-Bloch1}
\end{equation}
\begin{equation}
\frac{\partial W}{\partial t} = -2 i \kappa E_z (\rho - \rho^*) - \gamma_{\parallel} (W - W_\infty).
\label{MB-Bloch2}
\end{equation}

Similarly, we obtain the Maxwell-Bloch equations for the TE mode:
\begin{equation}
\left(\nabla_{x y}^{2}-\frac{n^2}{c^2} \frac{{\partial}^2}{\partial t^2} \right)H_z=2\beta\frac{n^2}{c^2} \frac{{\partial} H_z}{\partial t} 
-\frac{4\pi }{c}\frac{\partial}{\partial t}\left(\frac{\partial P_y}{\partial x}-\frac{\partial P_x}{\partial y}\right), 
\label{Eq2}
\end{equation}

\begin{equation}
\frac{1}{c} \frac{\partial}{\partial t} \left( \varepsilon E_x + 4 \pi P_x \right) = \frac{\partial H_z}{\partial y},
\label{MX-Ex-2}
\end{equation}
\begin{equation}
\frac{1}{c} \frac{\partial}{\partial t} \left( \varepsilon E_y + 4 \pi P_y \right) = - \frac{\partial H_z}{\partial x},
\label{MX-Ey-2}
\end{equation}
\begin{equation}
\PP = N \hbar \kappa (\rho + \rho^*) \ee,
\end{equation}
\begin{equation}
\frac{\partial \rho}{\partial t} = -(i \omega_0 + \gamma_\perp) \rho - i \kappa W E,
\label{MB-rho-TE}
\end{equation}
\begin{equation}
\frac{\partial W}{\partial t} = -2 i \kappa E (\rho - \rho^*) - \gamma_{\parallel} (W - W_\infty),
\label{MB-W-TE}
\end{equation}
where $E= \ee \cdot \EE$ is the electric field component along the polarization direction.

\subsection{Stationary Maxwell-Bloch equations for single mode lasing}


In this subsection, we demonstrate that the Maxwell-Bloch equations can be simplified to a set of time-independent equations when the laser operates in a single-mode regime.

\subsubsection{TM mode}

We consider the steady-state solution where the electric field oscillates at a single frequency $\omega$ for TM mode:  
\begin{equation}
E_z=E_s(x,y) e^{-i\omega t} + E_s^*(x,y) e^{i\omega t}.
\end{equation}
Accordingly, the microscopic polarization $\rho$ is assumed to take the following form: 
\begin{equation}
\rho=-i\rho_s(x,y) e^{-i\omega t}.
\end{equation}
By substituting these into Eqs.~(\ref{Eq1}) and (\ref{P-rho-relation}), we obtain the  following equation for the spatial profile $E_s(x,y)$:
\begin{equation}
 \left(\nabla^2_{xy} + n^2\frac{\omega^2}{c^2} \right)E_s=
 i4\pi\mu N\kappa\hbar \frac{\omega^2}{c^2}\rho_s 
-2i\frac{n^2}{c^2}\beta \omega E_s.
 \label{SMB-1}
\end{equation}
Substituting $E_s$ and $\rho$ into the Bloch equations $(\ref{MB-Bloch1})$ and $(\ref{MB-Bloch2})$   
and applying the rotating-wave approximation (RWA), we obtain the steady-state microscopic polarization $\rho_s$ and population inversion $W_s$
\begin{equation}
\rho_s=\dfrac{\kappa E_s}{\gamma_{\perp}+i(\omega_0-\omega)}W_s,
 \label{SMB-rho}
\end{equation}
and
\begin{equation}
W_s=\dfrac{W_{\infty} }
{
1+  
\frac{4\kappa^2}{ \gamma_{\parallel} }L(\omega) 
|E_s|^2  
}
, 
 \label{SMB-W}
\end{equation}
where $L(\omega)$ denotes the Lorentzian gain function 
\begin{equation}
L(\omega) \equiv 
\frac{\gamma_{\perp}}{(\omega-\omega_0)^2+\gamma_{\perp}^2}.
\end{equation}
Consequently, we finally obtain the equation for the stationary lasing of TM mode
\begin{equation}
\left(\nabla^2_{xy} + n^2 \frac{\omega^2}{c^2} \right)E_s= 
i \zeta \frac{\omega^2}{c^2}
L(\omega)
\left(
1+i\frac{\omega-\omega_0}{\gamma_{\perp}}
\right)
\dfrac{W_{\infty} }
{
1+ \frac{4\kappa^2}{ \gamma_{\parallel} }L(\omega) |E_s|^2  
}E_s -2i\beta \frac{n^2}{c^2}\omega E_s,  
\label{SMB-2}
\end{equation}
%
where 
\begin{equation}
\zeta \equiv 4\pi N\mu\kappa^2\hbar. 
\end{equation}
From Maxwell's equations, the boundary conditions for the electric field at the billiard boundary are given by
\begin{equation}
E_{s,in}=E_{s,out},
\label{TMboundary1}
\end{equation}
and
\begin{equation}
\frac{\partial E_{s,in} }{\partial n}=\frac{\partial E_{s,out} }{\partial n},
\label{TMboundary2}
\end{equation}
where $E_{s,in}$ and $E_{s,out}$ are the electric fields inside and outside 
the chaotic billiard laser, respectively, and $\partial/\partial n$ denotes 
the normal derivative at the billiard boundary. 
The first term on the right-hand side of Eq.~(\ref{SMB-2}) represents the effects of gain and saturation due to the lasing medium, while the second term represents absorption. Both terms vanish outside the cavity. In the exterior region, the radiation boundary condition (also referred to as the outgoing-wave boundary condition) must be satisfied because there are no reflecting structures outside the cavity.

Resonances derived from the linear Helmholtz equation in the absence of a lasing medium are characterized by complex eigenvalues. Precisely speaking, these are ``virtual states"—mathematical constructs rather than true stationary states. Historically, despite their inherent divergence at infinity, these resonances had long been the conventional choice for describing lasing states \cite{chang1996optical}.

A fundamental paradigm shift occurred when Eq.~(\ref{SMB-2}) was solved for a circular microcavity laser in \cite{harayama1999nonlinear}. This work provided a rigorous physical foundation by yielding real eigenvalues and physically consistent outgoing waves for the first time. It further challenged the traditional understanding of the total internal reflection condition, demonstrating that whispering gallery modes can achieve lasing even when their ray-orbit counterparts violate the critical angle. This new framework, which incorporates the lasing medium directly into the modal analysis, was subsequently advanced by Prof. A. Douglas Stone's group (Yale University) into a systematic theory for multimode lasing under weak mode interactions by neglecting modal interference terms called Steady-state Ab initio Laser Theory (SALT) \cite{tureci2006self, tureci2007theory, tureci2008strong}.

%
%

\subsubsection{TE mode}

Similarly to the TM mode, we assume that the electromagnetic fields and the microscopic polarization for the TE mode oscillate at a single frequency $\omega$ 
\begin{equation}
H_z=H_s(x,y)e^{-i\omega t} + H_s(x,y)^*e^{i\omega t},
\end{equation}
\begin{equation}
\EE = \{E_s(x,y)e^{-i\omega t} + E_s(x,y)^*e^{i\omega t} \}\e,
\end{equation}
\begin{equation}
\rho=-i \rho_s(x,y)e^{-i \omega t}.
\end{equation}
From the Bloch equations, we obtain the steady-state solutions  
\begin{equation}
\rho_s=\dfrac{\kappa E_s}{\gamma_{\perp}+i(\omega_0-\omega)}W_s,
 \label{SMB-rho-TE}
\end{equation}
and
\begin{equation}
W_s=\dfrac{W_{\infty} }
{
1+ 
\frac{4\kappa^2}{ \gamma_{\parallel} }L(\omega) |E_s|^2  
}.
 \label{SMB-W-TE}
\end{equation}
The resulting polarization is given by   
\begin{equation} 
\PP  =  N\kappa\hbar 
\left(
-i\rho_s e^{-i \omega t}+i\rho_s^* e^{i \omega t} 
\right) \e     
=
N\kappa ^2\hbar W_s
\left(
\frac{-ie^{-i \omega t}}{\gamma_{\perp}+i(\omega_0-\omega)} E_s +
\frac{ie^{i \omega t}}{\gamma_{\perp}-i(\omega_0-\omega)} E_s^*
\right) \e. 
\end{equation}
Consequently, by substituting these into Eq.~(\ref{Eq2}), we arrive at the governing equation for the stationary lasing of the TE mode:
\begin{equation}
\left(\nabla^2_{xy} + n^2\frac{\omega^2}{c^2} \right)H_s= 
\frac{\zeta}{\mu}
\frac{\omega}{c}
L(\omega) 
\left(
1+i\frac{\omega-\omega_0}{\gamma_{\perp}}
\right)
\left(
\frac{\partial}{\partial y} (W_s E_{x,s})-\frac{\partial}{\partial x} (W_s E_{y,s})
\right) 
-2i\beta \frac{n^2}{c^2}\omega H_s. 
 \label{SMB-TE}
\end{equation}
The boundary conditions for the magnetic field, derived from the Maxwell equations, are
\begin{equation}
H_{s,in}=H_{s,out},
\label{TEboundary1}
\end{equation}
and
\begin{equation}
\frac{1}{n_{in}^2}\frac{\partial H_{s,in} }{\partial n}=
\frac{1}{n_{out}^2}\frac{\partial H_{s,out} }{\partial n},
\label{TEboundary2}
\end{equation}
where $H_{s,in}$ and $H_{s,out}$ denote the magnetic fields inside and outside the 2D microcavity, respectively. Outside the cavity, the right-hand side of Eq.~(\ref{SMB-TE}) vanishes, and the radiation boundary condition must be satisfied.

\subsection{Linear stationary Maxwell-Bloch equations}
\label{Linear Laser}


When the pumping power is near the lasing threshold, the electric field intensity is sufficiently small that higher-order terms in the field can be neglected. Consequently, Eq.~(\ref{SMB-2}) reduces to the linearized stationary Maxwell-Bloch equation for the TM mode:
\begin{equation}
\left(\nabla^2_{xy} + n^2\frac{\omega^2}{c^2} \right)E_s=
i \zeta \frac{\omega^2}{c^2}
L(\omega) 
\left(
1+i\frac{\omega-\omega_0}{\gamma_{\perp}}
\right)
W_{\infty} E_s -2i\beta \frac{n^2}{c^2}\omega E_s.
\label{SMB-LINEAR-TM}
\end{equation}
The boundary conditions are given by Eqs.~(\ref{TMboundary1}) and (\ref{TMboundary2}). Outside the cavity, 
the right-hand side of Eq.~(\ref{SMB-LINEAR-TM}) vanishes, and the resonance wavefunction must satisfy the radiation boundary condition.

For the TE mode, by applying the RWA and considering the relationship between the electric and magnetic fields, the rotation term can be approximated near the threshold as  
\begin{equation}
\frac{\partial}{\partial y} (W_s E_{x,s})-\frac{\partial}{\partial x} (W_s E_{y,s}) \cong 
W_{\infty}\left(\frac{\partial E_{x,s}}{\partial y} -\frac{\partial E_{y,s}}{\partial x} \right) = i\omega W_{\infty} \frac{\mu}{c} H_s.
\end{equation}
Substituting this into Eq.~(\ref{Eq2}), we obtain
\begin{equation} 
\left(\nabla^2_{xy} + n^2 \frac{\omega^2}{c^2} \right)H_s=
i\zeta\frac{\omega^2}{c^2}
L(\omega)
\left(
1+i\frac{\omega-\omega_0}{\gamma_{\perp}}
\right) 
W_{\infty} H_s -2i\beta \frac{n^2}{c^2}\omega H_s.
\label{SMB-LINEAR-TE}
\end{equation}
The boundary conditions are given by Eqs.~(\ref{TEboundary1}) and (\ref{TEboundary2}).
The right-hand side of Eq.~(\ref{SMB-LINEAR-TE}) vanishes and the resonance wavefunction must satisfy the radiation boundary condition outside the cavity.

Consequently, the linearized Maxwell-Bloch equations for both TM and TE modes take precisely the same form:
\begin{equation}
\left(\nabla^2_{xy} + n^2 \frac{\omega^2}{c^2} \right)U_s=
i\zeta\frac{\omega^2}{c^2}
L(\omega)
\left(
1+i\frac{\omega-\omega_0}{\gamma_{\perp}}
\right) 
W_{\infty} U_s -2i\beta \frac{n^2}{c^2}\omega U_s,
\label{SMB-LINEAR}
\end{equation}
where $U_s$ represents $E_s$ and $H_s$ although the boundary conditions differ at the billiard boundary.

In the absence of external pumping ($W_\infty=0$), the solutions to Eq.~(\ref{SMB-LINEAR}) correspond to passive cavity resonances. These are quasi-stable states characterized by complex frequencies $\omega_j$. 
The real part $Re(\omega_j)$ represents the oscillation frequency, while the imaginary part $Im(\omega_j)$ (where $Im(\omega_j)<0$) 
accounts for the cavity loss or the decay rate of the resonance. 


As the external pumping power $W_{\infty}$ increases, the resonance frequency in the complex plane approaches the real axis, indicating a decrease in the effective decay rate. The shifted resonance frequency $\omega_j^{\prime}$ in the presence of linear gain can be approximated using the original passive resonance frequency $\omega_j$ as
\begin{equation}
{\omega_j^{\prime} }^2 \cong \omega_j^2 +i\zeta\frac{\omega_j^2}{n^2}
L({\operatorname{Re}\omega_j})
\left(1+
i\frac{\operatorname{Re}\omega_j-\omega_0}{\gamma_{\perp}}
\right)
W_{\infty} -2i\beta \omega_j .
\label{Shift}
\end{equation}
From the imaginary part of $\omega_j^{\prime}$, we derive the lasing condition for the mode $j$: 
\begin{equation}
\frac{\zeta}{2}\frac{\operatorname{Re}\omega_j}{n^2}
L({\operatorname{Re}\omega_j})
W_{\infty} > -\operatorname{Im}\omega_j +\beta.
\label{Threshold}
\end{equation}
At the lasing threshold, the gain exactly balances the cavity loss, causing the resonance to reach the real axis. From Eq.~(\ref{Shift}), the oscillation frequency at threshold is found to be
\begin{equation}
\omega_j^{\prime}\cong \frac{(-\operatorname{Im} \omega_j+\beta)\omega_0+
\gamma_{\perp}\operatorname{Re}\omega_j}{(-\operatorname{Im}\omega_j+\beta)+\gamma_{\perp}}.
\label{Shift-th}
\end{equation}
This result demonstrates frequency pulling, where the lasing frequency $\omega_j^{\prime}$ shifts toward the gain center 
$\omega_0$ as the cavity decay rate increases. 

It should be noted that for $W_{\infty}$ below the threshold, the solutions to the linear Maxwell-Bloch equation (\ref{SMB-LINEAR}) possess a negative imaginary part, which leads to unphysical divergence of the wavefunction at infinity. Conversely, above the threshold, the imaginary part becomes positive, and the wavefunctions no longer represent purely radiative states. Thus, the linear solutions are strictly appropriate for describing stationary lasing only at the exact threshold, where the frequency becomes real and the wavefunction satisfies the radiation boundary condition correctly. Consequently, the threshold solution of Eq.~(\ref{SMB-LINEAR}) provides a robust approximation of the stationary single-mode lasing state.

\subsection{Full nonlinear dynamical simulation of fully chaotic stadium lasers}

The linear approximation for single-mode lasing, as discussed in the previous subsection, is valid only in the weak-field regime. Notably, it fails to determine the intensity of the lasing mode due to the absence of saturation terms. Furthermore, as the pumping power increases, multiple resonant modes may acquire positive net gain. The subsequent asymptotic behavior of these modes—including mode competition and stabilization—can only be accurately captured through full nonlinear dynamical simulations of the Maxwell-Bloch equations.

To investigate the nonlinear dynamics of a fully chaotic laser, we consider a stadium billiard as the microcavity geometry. The stadium consists of two semicircles of radius $R=0.365~\mu\mathrm{m}$ connected by two parallel straight segments of length $L=2R=0.73~\mu\mathrm{m}$. 


The refractive indices are set to $n_{in}=3.3$ for the interior of the cavity, corresponding to typical semiconductor materials such as GaAs, 
and $n_{out}=1$ for the surrounding air. The transition frequency of the two-level atoms is $\omega_0 =2.2\times 10^{15}~\text{s}^{-1}$, which corresponds to a vacuum wavelength of approximately $\lambda_0\approx 857~\mathrm{nm}$ in the near-infrared range. 

The relaxation processes and other physical parameters used in our simulation are summarized as follows:
\begin{itemize}
\item Longitudinal relaxation rate: $\gamma_{\parallel}=6.6\times 10^{12}~\text{s}^{-1}$
\item Transversal relaxation rate: $\gamma_{\perp}=1.1\times 10^{15}~\text{s}^{-1}$
\item Internal loss coefficient: $8.8\times10^{12}~\text{s}^{-1}$
\item Atomic number density: $N=4.3\times 10^{18}~\text{cm}^{-3}$
\item Dipole coupling constant: $\kappa=1.1\times 10^{15}~\text{s}^{-1}$
\end{itemize}
Note that the value of $\gamma_{\perp}$ is significantly larger than $\gamma_{\parallel}$, indicating a fast dephasing limit typical of solid-state lasing media. 
The refractive index inside the billiard is $n_{\text{in}}=3.3$, while $n_{\text{out}}=1$. 

\begin{figure}[b]
\centering
\includegraphics[width=0.8\columnwidth]{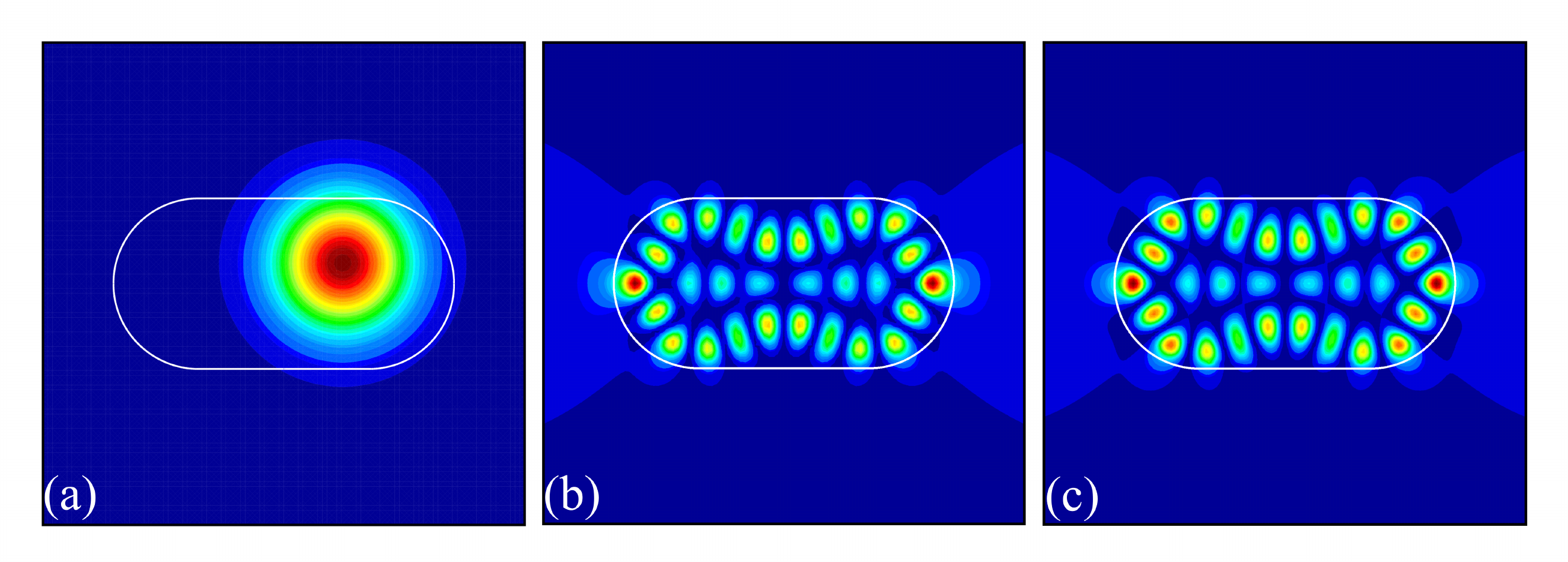}
\caption{
Intensity patterns of the fully chaotic stadium billiard lasers. 
(a) Initial profile of the electric field. The final stationary lasing state is found to be independent of the initial condition, demonstrating the stability of the oscillating mode. The white curve denotes the boundary of the stadium cavity.
(b) Spatial profile of the final stationary lasing state. The mode oscillates at the peak frequency identified in the power spectrum of Fig.~\ref{fig:SpecRes}(a) (corresponding to the resonance marked by the double circle in Fig.~\ref{fig:SpecRes}(b).)
(c) Eigenfunction of the quasi-stable resonance marked by the double circle in Fig.~\ref{fig:SpecRes}(b). 
The spatial distribution shows excellent agreement with the final stationary lasing state obtained from the dynamical Maxwell-Bloch simulation shown in (b).
Figures are adapted with permission from \cite{harayama2005theory}. Copyright 2005 by the American Physical Society.
}
\label{fig:wfs}
\end{figure}

We performed numerical simulations of the time evolution of the TM-mode electric field based on Eqs.~(\ref{MB-Bloch1})--(\ref{MB-Bloch2}), starting from an initial Gaussian wave packet as shown in Fig.~\ref{fig:wfs}(a). The dynamical evolution was calculated using the finite-difference time-domain (FDTD) method. To satisfy the radiation boundary condition at the edges of the computational domain, we implement a Perfectly Matched Layer (PML) to prevent unphysical reflections from the grid boundaries. The spatial grid size $\Delta x$ and time step $\Delta t$ are chosen to satisfy the Courant-Friedrichs-Lewy (CFL) condition, ensuring the stability and accuracy of the wave propagation. 

\begin{figure}[t]
\centering
\includegraphics[width=0.6\columnwidth]{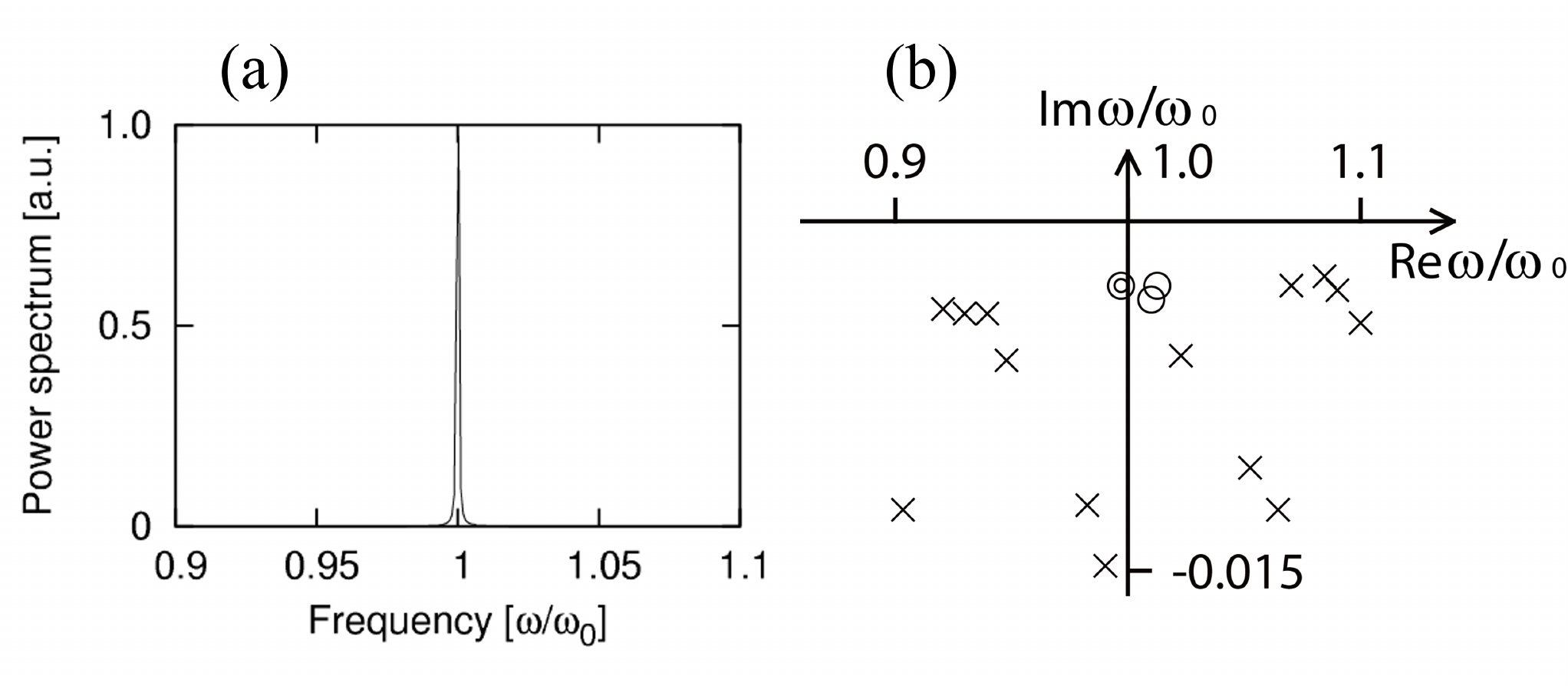}
\caption{
Spectrum and resonance distribution.
(a) Power spectrum calculated from the time evolution of the electric field during the stable oscillation regime. The dominant oscillation frequency shows excellent agreement with the real part of the passive resonance frequency indicated by the double circle in (b).
(b) Distribution of passive resonances for the microstadium cavity in the complex frequency plane. The double and single circles indicate the resonant modes with the highest and second-highest effective gains, respectively. The crosses represent modes that remain below the lasing threshold.
Figures are adapted with permission from \cite{harayama2005theory}. Copyright 2005 by the American Physical Society.
}
\label{fig:SpecRes}
\end{figure}

Initially, the total light intensity within the stadium is small; however, it grows exponentially due to the gain and eventually saturates, leading to a stable oscillation at a constant frequency. Figure~\ref{fig:SpecRes}(a) shows the power spectrum of the electric field in the post-saturation regime. The presence of a single sharp peak indicates that the system has reached a single-mode lasing state.

To identify this lasing mode, we separately calculated the passive cavity resonances (the ``cold cavity" modes) by solving the linear Maxwell-Helmholtz equation using the boundary element method (BEM) \cite{wiersig2003boundary}. The resulting distribution of resonances in the complex frequency plane is shown in Fig.~\ref{fig:SpecRes}(b). The modes denoted by the double and single circles satisfy the lasing condition [Eq.~(\ref{Threshold})], while the crosses represent sub-threshold modes.

The frequency of the sharp peak in the power spectrum (Fig.~\ref{fig:SpecRes}(a)) matches precisely the real part of the resonance frequency marked by the double circle in Fig.~\ref{fig:SpecRes}(b). This confirms that the laser action occurs on the specific resonance mode with the maximum effective gain.

Furthermore, the spatial profile of the final stationary lasing state obtained from the Maxwell-Bloch simulations (Fig.~\ref{fig:wfs}(b)) shows excellent agreement with the eigenfunction of the resonance mode marked by the double circle (Fig.~\ref{fig:wfs}(c)). This result demonstrates that, despite the presence of multiple potential modes, this particular resonance dominates the mode competition and achieves stable oscillation.

Interestingly, the spatial structure of the lasing mode in Fig.~\ref{fig:wfs}(b) does not appear to correspond to any simple closed ray trajectories (periodic orbits) of the stadium billiard. This highlights the complex wave-chaotic nature of the lasing modes in such geometries, where the mode structure is governed by the full wave dynamics rather than isolated classical paths.

Since this chapter is intended as an introduction to chaotic billiard lasers, in this subsection we have focused on the simplest case within nonlinear lasing theory: a fully chaotic billiard where a single resonance (cold-cavity mode) wins the mode competition to become a stationary lasing state. Consequently, we have precluded a detailed discussion of more profound nonlinear phenomena, such as spontaneous symmetry breaking of intensity patterns and universal single-mode lasing. In such regimes, stationary lasing states typically emerge as locked states of resonance wavefunctions, driven by complex mode-pulling and mode-pushing effects. Unlike conventional quantum/wave chaos, where wavefunctions do not interact, chaotic billiard lasers inherently involve such interactions. In this sense, chaotic billiard lasers provide a unique platform for investigating the double nonlinearity arising from the interplay between quantum/wave chaos and nonlinear laser dynamics 
\cite{Harayama2003, harayama2011two, sunada2013stable, sunada2016signature, harayama2017universal, you2022universal}.

\section{Summary and conclusions}

We have established a self-consistent theoretical and numerical framework to describe the physics of chaotic billiard lasers. 
By deriving the Maxwell-Bloch equations from first principles, we have bridged the gap between the linear theory of wave chaos 
and the nonlinear science of laser dynamics \cite{Harayama2003, harayama2011two}.

We have experimentally demonstrated chaos-assisted light emission in semiconductor chaotic billiard lasers. 
By employing a graded-index separate-confinement-heterostructure (GRIN-SCH) single-quantum-well structure 
and a specially designed inwardly recessed electrode contact window, we achieved the selective excitation of rectangular periodic ray orbits.
The near-field measurements using an IR-CCD camera provided direct evidence that the lasing mode is localized along the target rectangular orbit. 
Furthermore, the observation of two distinct emission points on each side of the cavity, which are spatially separated from the stable orbit, 
confirms the occurrence of dynamical tunneling from the stability islands into the surrounding chaotic sea. 
The resulting far-field patterns showed excellent agreement with numerical calculations, further validating that the emission is governed 
by the chaotic dynamics of the system.
These findings not only provide a fundamental understanding of chaos-assisted tunneling in open mesoscopic systems 
but also offer a novel approach to controlling the directional emission and modal properties of microcavity lasers. 
Our results pave the way for the design of high-performance chaotic lasers with tailored emission characteristics for photonic applications.

We have also elucidated the nonlinear Maxwell-Bloch formalism by providing a detailed derivation of the Maxwell-Bloch equations for 2D microcavities. 
We have demonstrated that for both TM and TE modes, the linearized equations near the threshold reduce to the same mathematical form, despite the fundamental differences in their respective boundary conditions. 
By analyzing the resonance movement in the complex frequency plane, we have identified the lasing threshold condition and analytically described the frequency pulling effect, showing how the interaction with the lasing medium forces the oscillation frequency toward the atomic transition center, consistent with established laser theory \cite{sargent1974laser}. 

Our full nonlinear FDTD simulations have confirmed that even in a complex, fully chaotic stadium geometry, the system can stabilize into a single-mode lasing state. The excellent spatial and spectral agreement between the dynamical results and the linear eigenfunctions validates the use of passive resonance analysis as a predictor of lasing modes near the threshold. Our analysis has revealed that the lasing mode in a fully chaotic stadium emerges from a complex process of mode competition governed by the global wave dynamics and the spatial distribution of the gain medium, rather than isolated classical periodic orbits.

In conclusion, chaotic billiard lasers provide a robust theoretical and experimental basis for understanding the complex interplay between laser physics and quantum/wave chaos, paving the way to exploit a new type of microcavity laser with novel functions.

Since this chapter is intended as an introduction to chaotic billiard lasers, the author has focused on providing detailed derivations of the fundamental equations, alongside explanations of key technical terms, concepts, and illustrative examples. To conclude, the author would like to outline several research branches related to chaotic billiard lasers for further exploration.

As previously pointed out, chaotic billiard lasers are characterized by "double nonlinearities" arising from the interplay between quantum/wave chaos and their interactions via an active lasing medium. The resulting lasing states are non-equilibrium steady states emerging from these complex nonlinear dynamics. However, these states are not always stationary; in some regimes, they exhibit permanent, irregular fluctuations. This behavior stems from the fact that lasing states obey the Maxwell-Bloch equations, which describe infinite-dimensional dissipative dynamics.

As the cavity size increases, the gain band encompasses numerous resonances whose interactions yield increasingly complex nonlinear dynamics. To investigate such regimes, the original Maxwell-Bloch equations become numerically inefficient because they must resolve the extremely fast optical carrier oscillations of the two-level atoms. Consequently, the complex dynamics of chaotic billiard lasers are often studied using the Schrödinger-Bloch model \cite{harayama2003stable, harayama2005theory, harayama2011two}. This model is derived by factoring out the rapid carrier oscillations, thereby eliminating the second-order time derivative in the Maxwell equation and transforming it into a Schrödinger-type equation.

Indeed, the rich and diverse behaviors exhibited by such nonlinear dynamics constitute the most fascinating aspect of chaotic billiard lasers. Although a comprehensive exploration of these exotic dynamical phenomena lies beyond the scope of this present chapter, the Schrödinger-Bloch framework serves as an indispensable tool for uncovering their underlying physical mechanisms.

In addition to brute-force simulations, mode-expansion analyses based on resonance wavefunctions have elucidated phenomena such as spontaneous symmetry breaking—driven by the locking of modes from different symmetry classes—and universal single-mode lasing \cite{Harayama2003, harayama2017universal}. These theoretical findings have been validated by experiments using semiconductor lasers \cite{choi2008alternate, sunada2013stable, sunada2016signature, you2022universal}, as well as by recent studies on laser dynamics in the vicinity of exceptional points \cite{matogawa2023nonlinear}.

In contrast to time-domain dynamical studies, Prof. A. Douglas Stone's group (Yale University) has developed the Steady-state Ab initio Laser Theory (SALT) \cite{tureci2006self, tureci2007theory, tureci2008strong, ge2008quantitative, tureci2009ab, ge2010steady, cerjan2011steady, cerjan2015steady}. This systematic framework determines steady-state multimode lasing solutions by assuming weak mode interactions and neglecting modal interference terms.

The problem of ray-wave correspondence in optical billiards differs significantly from that in quantum billiards; in the former, ray trajectories lose intensity upon each reflection at the boundary according to the Fresnel formulas. Consequently, the standard quantum ergodic theorem is not directly applicable to optical billiards. Numerical investigations, complemented by experiments on fully chaotic semiconductor billiard lasers, have confirmed that the ensemble average of resonance wavefunctions reproduces the distribution of ray trajectories weighted by the Fresnel coefficients \cite{ryu2006survival, shinohara2007signature, shinohara2008light, shinohara2009ray, choi2008dependence, harayama2015ray, you2022universal, ketzmerick2022chaotic, ketzmerick2025semiclassical}.

So far, we have focused on the fundamental physics of chaotic billiard lasers. Given their potential as advanced photonic devices, significant research has also been dedicated to their practical applications. A primary objective, initiated by the pioneering works \cite{Noeckel1997, Gmachl1998}, is the realization of low-threshold, highly directional light emission. Since then, this goal has been extensively pursued through a variety of design strategies \cite{PhysRevLett.88.094102, wiersig2006unidirectional, wiersig2008combining, yan2009directional, lee2011directional, kim2016designing, ryu2019optimization, park2019designing, park2021birefringent, lim2021robust, lee2023shape, Chen2016Highly, Jiang2017chaos, Schwefel:04, song2009chaotic, song2011highly, kurdoglyan2004unidirectional, yi2009lasing, kullig2025exceptional}.

Beyond directional emission, chaotic optical billiards and billiard lasers have been utilized in a wide range of other applications \cite{sunada2006sagnac, lee2008divergent, shim2008uncertainty, yang2010pump, song2012channeling, redding2012directional, redding2012local, ge2013extreme, song2013formation, redding2015low, sarma2015optical, sarma2015rotating, ge2015rotation, chen2019regular, chen2020chaos, doi:10.1126/science.abc2666, kim2023impact, kim2023spatiotemporal, kullig2020microstar, qian2021regulated, schrepfer2021dirac, wang2021direct, rodemund2024coupled, jiang2024coherent, sunada2019photonic, yamaguchi2023time, you2025nonlinear, ito2026expanding}.
Although space and time constraints preclude an exhaustive review of the diverse nonlinear phenomena and applications associated with chaotic billiard lasers, we hope this chapter serves as a solid foundation for readers to further explore the specialized literature on these emerging research areas.

\begin{ack}[Acknowledgments]

I initiated my theoretical studies of chaotic billiard lasers in collaboration with Dr.~Peter Davis (Telecognix) and Prof.~Kensuke Ikeda (Ritsumeikan University). Subsequently, I conducted experimental research alongside Prof.~Takehiro Fukushima (Okayama Prefectural University) and Dr.~Davis. I am deeply indebted to Prof.~Fukushima, from whom I learned all aspects of the fabrication and experimental techniques for microcavity lasers; indeed, nearly all the semiconductor lasers and experimental results presented in this chapter were provided by him.

I respectfully acknowledge their invaluable contributions, as well as Dr.~Bokuji Komiyama for his encouragement to pursue this field and the application of ring laser gyroscopes. I also wish to express my sincere gratitude to Prof.~Akira Shudo (Tokyo Metropolitan University) for introducing me to the field of quantum chaos and for his guidance in my research on this area.

Special gratitude is extended to Prof.~Susumu Shinohara (Komatsu University) and Prof.~Satoshi Sunada (Kanazawa University) for their continuous cooperation and support; the numerical calculations presented in this chapter were largely performed by them. Prof.~Sunada has extended the research on chaotic billiard lasers to include innovative applications in machine learning \cite{sunada2019photonic, yamaguchi2023time, you2025nonlinear, ito2026expanding}. His pioneering work in this direction has attracted significant attention.

I would also like to express my sincere gratitude to the members of my laboratory for their dedicated support. In particular, I am grateful to Dr.~Mengyu You, Mr.~Yoshikazu Kuribayashi, Ms.~Maika Matogawa, Mr.~Kazumune Matsumoto, and Ms.~Moka Kurita for their significant contributions. My thanks also extend to Mr.~Shun'ya Sekiguchi, Mr.~Yuta Kawashima, Mr.~Yuichiro Suzuki, and all the other former members who have supported my research over the years.

I also wish to thank Profs.~Martina Hentschel, Roland Ketzmerick, Arnd B\"{a}cker, Jan Wiersig, Hans-J\"{u}rgen St\"{o}ckmann, Muhan Choi, Jung-Wan Ryu, A. Douglas Stone, Hui Cao, Qinghai Song, Hakan E. T\"{u}reci, Harald G. L. Schwefel, Li Ge, Stefan Rotter, Claire F. Gmachl, Henning Schomerus, Pierre Gaspard, Rainer Klages, Felipe Barra, Bertrand Georgeot, Barbara Dietz, Melanie Lebental, Stefan Bittner, Kyungwon An, Sang-Bum Lee, Chil-Min Kim, Sang Wook Kim, and Yun-Feng Xiao, as well as the late Profs.~Shuichi Tasaki and Soo-Young Lee, for many fruitful discussions. 

Finally, I would like to express my deepest gratitude to Professor Karol Życzkowski and Professor Sven Gnutzmann for providing the opportunity to contribute to the quantum chaos section of this Encyclopedia.

This work was supported by JSPS KAKENHI Grant Number JP22H05198 and Waseda University (Grant Numbers 2025Q-021 and 2025C-739).
\end{ack}


\bibliographystyle{JHEP}%
\bibliography{ref.bib}

\end{document}